\documentclass[]{bmcart}

\usepackage{amsthm,amsmath}
\usepackage{amssymb}
\RequirePackage{natbib}
\usepackage[utf8]{inputenc} 
\usepackage{graphicx}
\usepackage{xspace}
\usepackage{tikz}
\usetikzlibrary{arrows,positioning}
\usepackage[caption=false]{subfig}
\usepackage[resetfonts]{cmap}
\usepackage{fancyvrb}
\begin{VerbatimOut}{ot1.cmap}
	/CIDInit /ProcSet findresource begin
	12 dict begin
	begincmap
	/CIDSystemInfo
	<< /Registry (TeX)
	/Ordering (OT1)
	/Supplement 0
	>> def
	/CMapName /TeX-OT1-0 def
	/CMapType 2 def
	1 begincodespacerange
	<00> <7F>
	endcodespacerange
	8 beginbfrange
	<00> <01> <0000>
	<09> <0A> <0000>
	<23> <26> <0000>
	<28> <3B> <0000>
	<3F> <5B> <0000>
	<5D> <5E> <0000>
	<61> <7A> <0000>
	<7B> <7C> <0000>
	endbfrange
	40 beginbfchar
	<02> <0000>
	<03> <0000>
	<04> <0000>
	<05> <0000>
	<06> <0000>
	<07> <0000>
	<08> <0000>
	<0B> <0000>
	<0C> <0000>
	<0D> <0000>
	<0E> <0000>
	<0F> <0000>
	<10> <0000>
	<11> <0000>
	<12> <0000>
	<13> <0000>
	<14> <0000>
	<15> <0000>
	<16> <0000>
	<17> <0000>
	<18> <0000>
	<19> <0000>
	<1A> <0000>
	<1B> <0000>
	<1C> <0000>
	<1D> <0000>
	<1E> <0000>
	<1F> <0000>
	<21> <0000>
	<22> <0000>
	<27> <0000>
	<3C> <0000>
	<3D> <0000>
	<3E> <0000>
	<5C> <0000>
	<5F> <0000>
	<60> <0000>
	<7D> <0000>
	<7E> <0000>
	<7F> <0000>
	endbfchar
	endcmap
	CMapName currentdict /CMap defineresource pop
	end
	end
\end{VerbatimOut}

\startlocaldefs

\newif\ifdraft
\newif\ifproof
\drafttrue
\prooftrue
\draftfalse

\def\num@def#1{\def\num@name{#1}\edef\num@str{\string #1}%
	\expandafter\ifx\csname\num@str{*}\endcsname
	\relax\num@start\fi\futurelet\next\num@d@f}%

\def\num@d@f{\ifx\next*\let\@tempa\num@defA
	\else\ifx\next\bgroup\let\@tempa\num@defB
	\else\let\@tempa\num@defC \fi\fi\@tempa}%

\newcount\num@count
\def\N{\number\num@count}%

\def\num@defA*{\def\num@{*}\num@def@}%
\def\num@defB#1{\let\num@=\N\num@count= #1 \num@def@}%
\def\num@defC{\let\num@=\N\afterassignment\num@def@\num@count= }%

\def\num@def@{\expandafter\num@define\csname\num@str{\num@}\endcsname}%

\def\num#1#2{\let\num@define=#1\num@def#2}%

\def\num@start{%
	\expandafter\ifx\num@name\@undefined\let\next\num@st@rt
	\else\def\next{\errmessage{numdef: \num@str\space is already defined}}%
	\fi\next}%

\def\num@st@rt{\expandafter\def\csname\num@str{*}\endcsname
	{\errmessage{numdef: \num@str\N\space is not defined}}%
	\expandafter\num@use\num@name}%

\def\num@use#1{\def#1{\edef\num@str{\string #1}\futurelet\@tempa\num@us@}}
\def\num@us@{%
	\ifx\@tempa\bgroup\let\next\num@useA\else\let\next\num@useB\fi\next}

\def\num@useA#1{\num@count= #1 \num@use@}%
\def\num@useB{\afterassignment\num@use@\num@count= }%
\def\num@use@{\expandafter\ifx\csname\num@str{\N}\endcsname\relax
	\def\num@{*}\else\let\num@=\N\fi\csname\num@str{\num@}\endcsname}%

\usepackage{array}
\ifdraft \newcolumntype{H}{l} \else\newcolumntype{H}{>{\setbox0=\hbox\bgroup}c<{\egroup}@{}}\fi
\newcolumntype{Z}{>{\setbox0=\hbox\bgroup}c<{\egroup}@{\hspace*{-\tabcolsep}}}

\definecolor{OliveGreen}{rgb}{0,0.3,0}
\definecolor{RedOrange}{rgb}{0.8,0.2,0.1}
\definecolor{Blu}{rgb}{0.3,0.3,0.8}

\usepackage{ulem}
\newcommand{\acom}[1]{$\{${\color{RedOrange}#1}$\}$}

\newcommand{\ecom}[1]{$\{${\color{Blu}#1}$\}$}

\newcommand{\gcom}[1]{$\{${\color{OliveGreen}#1}$\}$}

\ifdraft
\else
	\renewcommand{\acom}[1]{\ }
	
	\renewcommand{\ecom}[1]{\ }
	
	\renewcommand{\gcom}[1]{\ }
	
\fi

\long\def\/*#1*/ {}

\newcommand{\OpT}{{\Delta_T}}

\newcommand{\M}{M}
\newcommand{\period}{P}
\newcommand{\T}{T} 
\newcommand{\fu}{\mu Pa}

\newcommand{\irange}{$R_{\textrm{inter}}$\xspace}
\newcommand{\fref}[1]{fig.~\ref{#1}}
\newcommand{\tref}[1]{tab.~\ref{#1}}
\newcommand{\surv}{$\Psi(t)$\xspace}
\newcommand{\nagent}{8890\xspace}

\newcommand{\ume}{Unit of measure} 
\newcommand{\avd}[1]{\overline{\Delta #1}_p} 
\newcommand{\cm}[1]{}
\newcommand{\RA}[1]{T_{#1}}

\newcommand{\Int}{\mathrm{I}}
\newcommand{\RAF}{\mathrm{RAF}}
\newcommand{\RAS}{\mathrm{RAS}}
\newcommand{\FAK}{\mathrm{FAK}}
\newcommand{\ERK}{\mathrm{ERK}}
\newcommand{\MEK}{\mathrm{MEK}}
\newcommand{\RUNX}{\mathrm{RUNX2}}
\newcommand{\Other}{\mathrm{OTHER}}
\newcommand{\OSXmon}{\mathrm{OSX}_\mathrm{mon}}
\newcommand{\OSXmul}{\mathrm{OSX}_\mathrm{mul}}
\newcommand{\DNA}{\mathrm{DNA}}
\newcommand{\AP}{\mathrm{AP1}}
\newcommand{\OPN}{\mathrm{OPN}}
\newcommand{\OCN}{\mathrm{OCN}}
\newcommand{\ALP}{\mathrm{ALP}}
\newcommand{\BSP}{\mathrm{BSP}}
\newcommand{\Rib}{\mathrm{Rib}}
\newcommand{\Ves}{\mathrm{Vesc}}
\newcommand{\RNA}{\mathrm{RNA}}
\newcommand{\MAT}{\mathrm{MAT}}
\newcommand{\comp}{\textrm{c}}
\newcommand{\ncomp}{\textrm{n comp}}
\newcommand{\aval}{\textrm{av}}
\newcommand{\naval}{\textrm{n av}}
\newcommand{\act}{\mathrm{act}}
\newcommand{\fact}{\textrm{f act}}
\newcommand{\nact}{\mathrm{d}}
\newcommand{\Cell}{\textrm{Osteoblast cell}}
\num\newcommand{\s00}{\Int_\nact}
\num\newcommand{\s04}{\Int_\act}
\num\newcommand{\s01}{\Int_\fact}
\num\newcommand{\s02}{\Int_\fact+\FAK_\act} 
\num\newcommand{\s10}{\FAK_\act} 
\num\newcommand{\s11}{\Int_\fact+\FAK_\act} 
\num\newcommand{\s12}{\Int_\fact+\FAK_\act+\RAS_\nact} 
\num\newcommand{\s13}{\FAK_\nact}
\num\newcommand{\s20}{\RAS_\nact}
\num\newcommand{\s21}{\Int_\fact+\FAK_\act+\RAS_\nact} 
\num\newcommand{\s22}{\RAS_\act}
\num\newcommand{\s23}{\RAS_\act+\RAF_\nact} 
\num\newcommand{\s30}{\RAF_\nact}
\num\newcommand{\s31}{\RAS_\act+\RAF_\nact} 
\num\newcommand{\s32}{\RAF_\act}
\num\newcommand{\s33}{\RAF_\act+\MEK_{\nact}} 
\num\newcommand{\s40}{\MEK_\nact}
\num\newcommand{\s41}{\RAF_\act+\MEK_\nact}
\num\newcommand{\s42}{\MEK_\act}
\num\newcommand{\s43}{\MEK_\act+\ERK_\nact} 
\num\newcommand{\s50}{\ERK_\nact}
\num\newcommand{\s51}{\MEK_\act+\ERK_\nact} 
\num\newcommand{\s52}{\ERK_\act}
\num\newcommand{\s53}{\ERK_\act+\RUNX_\nact}
\num\newcommand{\s54}{\ERK_\act+\Other}
\num\newcommand{\s60}{\RUNX_\nact}
\num\newcommand{\s61}{\ERK_\act+\RUNX_\nact}
\num\newcommand{\s62}{\RUNX_\act}
\num\newcommand{\s63}{\RUNX_\act+\DNA}
\num\newcommand{\s70}{{\OSXmon}} 
\num\newcommand{\s700}{\RUNX_\act+\OSXmon}
\num\newcommand{\s71}{\OSXmon+\DNA}
\num\newcommand{\s710}{\RUNX_\act+\OSXmon+\DNA}
\num\newcommand{\s711}{\RUNX_\act+\OSXmon+\DNA+\AP}
\num\newcommand{\s72}{\OSXmul}
\num\newcommand{\s720}{\RUNX_\act+\OSXmul}
\num\newcommand{\s73}{\OSXmul+\DNA}
\num\newcommand{\s730}{\RUNX_\act+\OSXmul+\DNA}
\num\newcommand{\s80}{\AP}
\num\newcommand{\s81}{\RUNX_\act+\OSXmon+\DNA+\AP}
\num\newcommand{\s90}{\OPN_\RNA}
\num\newcommand{\s91}{\Rib_\aval+\OPN_\RNA}
\num\newcommand{\s92}{\Rib_\naval+\OPN_\RNA}
\num\newcommand{\s100}{\OCN_\RNA}
\num\newcommand{\s101}{\Rib_\aval+\OCN_\RNA}
\num\newcommand{\s102}{\Rib_\naval+\OCN_\RNA}
\num\newcommand{\s110}{\ALP_\RNA}
\num\newcommand{\s111}{\Rib_\aval+\ALP_\RNA}
\num\newcommand{\s112}{\Rib_\naval+\ALP_\RNA}
\num\newcommand{\s120}{\BSP_\RNA}
\num\newcommand{\s121}{\Rib_\aval+\BSP_\RNA}
\num\newcommand{\s122}{\Rib_\naval+\BSP_\RNA}
\num\newcommand{\s130}{\Rib_\ncomp}
\num\newcommand{\s131}{\Rib_\ncomp+\RNA}
\num\newcommand{\s140}{\Rib_\comp}
\num\newcommand{\s141}{\Rib_\comp+\RNA}
\num\newcommand{\s150}{\MAT+\Ves}
\num
\num
\num
\num
\num\newcommand{\s1000}{\Cell}
\num\newcommand{\RADP10}{$\RA{\s41 \rightarrow \s42}$}
\num\newcommand{\RADP11}{$\RA{\s43 \rightarrow \s42}$}
\num\newcommand{\RADP12}{$\RA{\s42 \rightarrow \s40}$}
\num\newcommand{\RADP13}{$\RA{\s51 \rightarrow \s52}$}
\num\newcommand{\RADP14}{$\RA{\s52 \rightarrow \s50}$}
\num\newcommand{\RADP15}{$\RA{\s53 \rightarrow \s52}$}

\endlocaldefs

\begin{document}

\begin{frontmatter}

\begin{fmbox}
\ifproof \else \dochead{Research}\fi


\title{ABM of osteoblast's mechanotransduction pathway: time patterns of critical events}


\ifproof
\author[
	addressref={shefOM,insigneo},
	email={g.ascolani@sheffield.ac.uk}   
]{\inits{GA}\fnm{Gianluca\vspace{1cm}} \snm{Ascolani}}
\author[
addressref={shefOM},
email={t.skerry@sheffield.ac.uk}
]{\inits{TS}\fnm{Timothy M.} \snm{Skerry}}
\author[
addressref={insigneo,shefEng},
email={}
]{\inits{DL}\fnm{Damien} \snm{Lacroix}}
\author[
addressref={shefOM,insigneo},
email={e.dallara@sheffield.ac.uk}
]{\inits{ED}\fnm{Enrico} \snm{Dall'Ara}}
\author[
addressref={shefOM,insigneo},
corref={shefOM},        
email={aban.shuaib@sheffield.ac.uk}
]{\inits{AS}\fnm{Aban} \snm{Shuaib}}

\setattribute{corref}           {text} {}
\setattribute{authorinfo}       {text} {}

\makeatletter
\setvaluelist{bmcsymbol}{{},\textdagger,\^{}}
\def\auto@set@thanksbox{}

\def\printcorrtext#1{%
	\corref@thanksmark{#1}%
	\corref@text%
}

\def\auto@set@thanksbox{%
		\@ifundefined{corr@author@id}{}{\printcorrtext{\corr@author@id}}%
		\@ifundefined{corref@list}{}{\@for\address@id:=\corref@list\do{}\par}%
	}

\makeatother

\else
\author[
	addressref={shefOM,insigneo},
	email={g.ascolani@sheffield.ac.uk}   
]{\inits{GA}\fnm{Gianluca} \snm{Ascolani}}
\author[
addressref={shefOM},
email={t.skerry@sheffield.ac.uk}
]{\inits{TS}\fnm{Timothy M.} \snm{Skerry}}
\author[
addressref={insigneo,shefEng},
email={}
]{\inits{DL}\fnm{Damien} \snm{Lacroix}}
\author[
addressref={shefOM,insigneo},
email={e.dallara@sheffield.ac.uk}
]{\inits{ED}\fnm{Enrico} \snm{Dall'Ara}}
\author[
addressref={shefOM,insigneo},
corref={shefOM},        
email={aban.shuaib@sheffield.ac.uk}
]{\inits{AS}\fnm{Aban} \snm{Shuaib}}
\fi%



\address[id=shefOM]{
  \orgname{Department of Oncology \& Metabolism, University of Sheffield}, 
  \city{Sheffield},                              
  \cny{UK}                                    
}
\address[id=insigneo]{
	\orgname{Insigneo Intitute for in \textit{silico} medicine, University of Sheffield}, 
	\city{Sheffield},                              
	\cny{UK}                                    
}
\address[id=shefEng]{
	\orgname{Department of Mechanical Engineering, University of Sheffield}, 
	\city{Sheffield},                              
	\cny{UK}                                    
}



\begin{artnotes}
\note[id=n1]{Equal contributor}  
\end{artnotes}

\ifproof \else\end{fmbox}\fi 


\begin{abstractbox}
\begin{abstract} 
\parttitle{Background} 
Mechanotransduction in bone cells 
plays a pivotal role in
osteoblast differentiation and bone remodelling. 
Mechanotransduction provides the link between modulation of the extracellular matrix and intracellular actions.
By controlling the balance between the intracellular and extracellular domains, the mechanotransduction process determines optimal functionality of the skeletal dynamics, and it is one of the possible causes of osteophatological diseases.
\parttitle{Results} 
Mechanotransduction in a single osteoblast under external mechanical perturbations has been modelled in the agent based framework to reproduce the dynamics of the stochastic reaction diffusion process among molecules in the cytoplasm, nuclear and extracellular domains. The amount of molecules and fluctuations of each molecular class has been analysed in terms of recurrences of critical events. A numerical approach has been developed to invert subordination processes and to extract the directing processes from the molecular signals in order to derive the distribution of recurrence of events.
\parttitle{Conclusions} 
We observed large fluctuations enclose information hidden in the noise which is beyond the dynamic variations of molecular  baselines. 
Studying the system under different parametric conditions and stimuli, the results have shown that the waiting time distributions of each molecule are \/* informative*/ a signature of the system's dynamics. 
The behaviours of the molecular waiting times change with the changing of parameters presenting 
the same variation of patterns for similar interacting molecules and identifying specific alterations \/*properties*/ for key molecules in the mechanotransduction pathway. 



\end{abstract}




\end{abstractbox}

\ifproof\end{fmbox}\fi

\end{frontmatter}




\ifproof\pagebreak\fi
\section*{Background}

Mechanotransduction drives the orchestration between bone remodelling, the process of bone adaptation over time, and the activity of the bone cells: osteoblasts, osteoclasts and osteocytes \cite{yavropoulou2016,chen}. 
Transmission of mechanical stimulations through extracellular matrix (ECM) medium deep inside the cells affecting the intracellular signalling, both in the cytoplasm and in the nucleus is a phenomenon which, despite significant progress in recent years, has been found more complex than it has been thought \cite{yap,fried,harris,alonso,martino}.
Studying the deviations from the coordinated balance of mechanotransduction signalling in osteoblasts, which can cause osteodegenerative diseases such as osteoporosis, osteopetrosis, and osteoarthritis, is fundamental \cite{humphrey,vincent,papachroni}.

Effects of mechanical loads propagate through the bone structure, stimulating both the osteocytes with their dendritic-like processes and the lining cells on the bone apposition surface. The production of osteoapatite and factors involved in the conformation of the ECM  is a continuous dynamic process occurring in both cases of fracture repair and tissue deformation.
The mechanical stimulations propagated through the tissue are sensed by osteocytes and osteblasts via mechanoreceptors such as integrins \cite{bonewald}. 

Integrins play a pivotal role in migration as well as in osteogenic differentiation and osteoblast activation \cite{yap,fried,oh_chi,charras}. \/*Integrins play a pivotal role in migration as well as in the differentiation and maturation of osteoblasts \cite{yap,fried,oh_chi,charras}.*/ 
The extracellular domain of integrins provides anchorage and interaction interface with the ECM, while ensembles of integrins, on the intracellular domain, are molecular sensors accumulating the effects of the external focal adhesion forces.
One of the main aspects of integrin mediated signals is the cumulation of structural tension in the protein, and its consequent activation is antagonized by a relaxation process \cite{kim,zhu}. Thus, transduction can be strongly prolonged if the external force does not produce adequate deformations.

Stretching and compression of non homogeneous tissue locally depends on the mechanical properties of the extracellular milieu. Considering the limited diffusion of large non soluble molecules in dense ECM, the local properties of the latter are mainly affected by cell secretion of factors which accumulate in the surrounding volume. In turn, variation of the ECM stiffness changes how a cell senses the forces. 
Therefore, the signal conveyed by integrins affects bone tissue at different scales \cite{buo,thiel,sun}. 

Mechanotransduction has been investigated with experimental \cite{wang,humphrey,muhamed} and computational  \cite{shams,preziosi} approaches; nevertheless, many aspects of this interaction in bone cells have not yet been explored in large part due to the lack of accurate tools to study molecular events in live cells. 
Many of the in vitro setups are limited to 2D distributions of cells with reduced density of surrounding ECM in order to facilitate stimulations and observations of the system. Measurements based on traction force microscopy are affected by low spatial resolution and constraints in the orientation of applied forces \cite{piel}. 

Molecular tension sensors experiments are useful for observation of downstream chemical signalling concurrent with fluorescence signals. However, spectrum limitations can be challenging when downstream signalling is being explored. Furthermore, Foster resonance energy transfer (FRET) experiments often limit the number of interactions, the range of extensions and forces that can be explored \cite{jurchenko}. 
In silico Molecular Dynamics models are limited to a simulated time which is in the order of 100 nanoseconds \cite{kiseleva}, while continuous approaches lack the possibility to consider discrete and highly heterogeneous systems that can form local recurrent structures among the molecules \cite{peng}.   

An agent based approach can overcome some of the above mentioned limitations of present in vitro and in silico approaches. 
The advantage of using an agent based model (ABM) consists in generating a complex dynamics based on simple rules for the physical interactions that would not be easily obtainable in a continuous representation of the system and would be  experimentally difficult to control.
ABMs are suitable to mimic 3D cellular systems where the dynamics of the signalling and the occurrence of events  inside a cell can be tracked at the molecular scale \cite{dong} Furthermore, the heterogeneity of molecular populations  responding to directional and time modulated stimuli can be considered \cite{ausk} Additionally, these approaches can be adopted to address the effects of non homogeneity in the plasm and nucleus of the cell far from the condition of well mixed systems assumed in the Gillespie class of algorithms \cite{zhang}. In contrast to MD and ODEs/PDEs methods, the coarsening of molecules to single diffusing interacting agents accounts for intracellular stochastic effects and yet, run large simulations for periods of time corresponding to one day \cite{bonchev, peng}.

The aim of this study was to use an ABM model of the osteoblast to study the effect of external mechanical stimuli to the local molecular events within this cell. 
\section*{Implementation}
We used an ABM to mimic the dynamics in the mechanotransduction pathway of a single osteoblast during the deposition of 
ECM factors 
under external mechanical perturbations
in order to analyse fluctuations of molecular signals and to extract time patterns of critical events characteristic of the process dynamics.
In the ABM, each agent represents a specific molecule, which has a relevant role in the process of ECM formation
and is included in the mechanotransduction network (see \fref{fig_mechnet}). The agents move following a constrained Brownian motion in one of three geometrical compartments – the cell nucleus, the cytoplasm, and the extracellular region surrounding the cell (orange borders in \fref{fig_mechnet}).  
When the distance between two agents is shorter than the interaction range \irange, a set of rules depending on the internal state of the agents is applied to each agent involved in the interaction.
Similarly, single agents can trigger rules due to their internal clock, state or information exchanged.
The actions resulting from the application of rules triggered by an agent may involve:
\begin{enumerate}
	\item 
	the changing of its own state, 
	\item the self-destruction, 
	\item the production of information causing: 
	\begin{enumerate}
		\item 
		other existing agents to change their states, as well as 
		\item 
		the creation of new agents, or
		\item 
		triggering other actions. 
	\end{enumerate}
\end{enumerate}

In terms of agents, the reactions between two molecules generating a molecular complex are two types. The first requires the change of state of one agent and the disappearance of the other; the second consists in the disappearance of both agents involved and the appearance of a new agent. Reactions referring to a transfer of a molecular messenger are obtained by the change of the states of the agents involved. Dissociations in two or more molecules of complexes regulated by the delay and molecular persistence in a given state have two ways. One way is the disappearance of the original agent and the appearance of as many agents as the number of dissociated molecules. The other way consists of  creating as many agents as the number of dissociated molecules minus one and the change of state of the original agent.

Among the molecules simulated there are 
proteins (e.g. integrins), ribosomes, mRNA and vesicles. For the sake of simplicity,  other elements like amino acids, organelles, molecules and complexes not included in \fref{fig_mechnet} are neglected. 
The rules we used in the proposed model  are based on the affinity of the molecules, 
and they are derived by the chemical reactions sketched in the molecular network diagram. 

The updating scheme adopted  \cite{gizhub} requires that, at each iteration, all the molecules are swept and checked for possible rules to be applied. 
The types of physical interaction driving the model dynamics are two: 1) binary binding of molecules  based on one molecule querying all the affine available molecules in the sphere with radius \irange and then requesting the interaction with the nearest one; 2) degradation, relaxation and unstable complex dissociation activated at the expiration of an internal timer set to a random time tossed from a probability function  \surv, when the complex changes state and decreased of 1 unit of time at each iteration.

In the simulated ABM, among various signals, we generated the cumulated number of agents partitioned by their internal biological \/*meaningful*/ state variables 
potentially representing  experimental distinguishable molecular subpopulations (e.g. active/inactive and bound/unbound).

Many of the simulated molecules are present, at least, in two complementary isoforms: active and inactive. In macroscopic steady state conditions, when one isoform produces a smooth signal close to zero axis interrupted by positive spikes, the other generates a negative fluctuation around a positive nominative value. Indeed, the process becomes immediately much more complex when there are not only active and inactive states, but also interactions with other molecular species [4,5]. 
Comparing multiple simulations, it can be easily spotted that spikes between different repetitions are not synchronized, and the time intervals between spikes of the same signal are not constant. These fluctuations, typical of random processes, can be difficultly associated to specific reactions when in presence of a heterogeneous population of agents undergoing different interactions. Nonetheless, formation of patterns and information of the complexity of a non-ergodic system can be extracted.

The former type of isoforms suggests studying the signals composed mainly of positive spikes emerging from zero in terms of stochastic analysis of 
large fluctuations and noise also when in presence of external perturbations. Nonetheless, the generalization of the analysis to other molecules requires us to cope with the nominal values which are different from zero and change in time during the simulations due to variation of the equilibrium condition. For these reasons, we have developed a method to deal with any types of signals produced by this ABM.

\subsection*{Signals’ transformation}

The agents generated by the ABM contain time dependent information relative to their position, velocity, belonging compartment, molecular activation states, bound state and an internal clock. If each agent has $n$ variables, and there are $S$ agents, then the state of the system can be represented in a $n\times S$ dimensional phase space. In order to reduce the dimensionality of the system’s phase space, in this study we neglected all the information relative to the position and velocity 
and we considered only the molecular activation state, the bound state and, when biologically relevant, the cellular compartment. To be precise, these quantities are categorical; therefore, we can label them with discrete indices.
The signals analysed are the sum of molecules with the same labelling. Let us call each of these quantities state variables which have values $y_i(t)$, where $i$ is the index for the specific state variable and $y$ is the value dependent on the time $t$. Given the stochastic origin of the simulations, the resulting time series  are fluctuating signals, which can be decomposed in two components: the trend, $Y_i(t)$, and the fluctuations around the trend, $f_i(t)$. We introduce a method to process the signal in order to obtain a positive defined time series of the noisy component
by using the time average, a filter that removes the random fluctuations faster then a time lapse $\OpT$:

$$\overline{y_i}(t)= \widehat{A}_\OpT y_i(t) \xrightarrow{\OpT\rightarrow\infty} \overline{Y_i}(t),$$
where the over bar represents the time averaged signal and the operator  is the moving average operator:
$$\widehat{A}_T=\frac{1}{\OpT}\int_{t}^{t+\OpT}(\cdot)\ dt^\prime.$$


This approach allowed us to sequentially cut out fluctuations at different and specific time scales and to analyse the results also for a system in an out of equilibrium condition perturbed by external mechanical signals sensed by the integrins on the membrane of the osteoblast.

The periodic step-wise external perturbation changes the equilibrium state accordingly to the frequency, and, for fast variations compared to the response of the molecules, the system remains out of equilibrium for the whole simulation while the molecules try to reach the equilibrium condition. The average of the signal, being non constant in time, is one of the reasons why the system is not ergodic. Therefore, we first remove the trend from each signals in order to obtain 
the fluctuations around the trend.


The choice of using a time average operator instead of the standard ensemble average has been dictated by biological motivations.  
Mathematically, an ensemble average is intended as an average among different regions which are statistically independent, while, biologically speaking, it could be reinterpreted as an average of non communicating parts of the system. Therefore, there are no real means for the molecules
with a finite interaction range to perceive a global average signal.
On the contrary, 
a time average represents the  slow component of the dynamics cumulated into the 
disposition of the molecules which can affect the relaxation of large deviations.
Therefore, for a biological complex system, a time average operator seams physically more meaningful
for comparing fluctuations against the trend of the system.



\subsection*{Subordination theory and detection of events}
In this work, we propose a numerical method to disentangle random processes called subordinated processes \cite{sokolov} into their respective parent process and directing process in order to analyse the latter in terms of patterns of recurrence of events \cite{mainardiex}.   
Let us show how to generate a subordinated process \cite{montroll,feller2,mainardi1,mainardi2,gringo, ascolani}.
First, we generate a leading process, which can be deterministic or derived from a random distribution $X$ so that a trajectory $x$  
has values 
$x(t^\star_i)$
at discrete positive times $t^\star_i$ with  $t^\star_i-t^\star_{i-1}=\Delta t^\star>0\quad\forall\ i\in\mathbb{N}$. If we rescale the time such that  $\Delta t^\star=1$ unit of time, then $t^\star_i=i$ and the parent process is described by 
$x(t^\star_i)=x(i)$
where the steps 
$x(t^\star_i)-x(t^\star_{i-1})$
are tossed from $X$.
Second, we generate in a similar way a directing process  $T$ from a random distribution with positive support such that the tossed $\tau_i$ defines a monotonically increasing function
$t(t^\star_i)=\sum_{j=0}^i\tau_j$. 
The subordinated function is defined as the trajectory $x(t)$. 
Generally $t^\star$ is called the natural time and  $t$ is called the physical time, which is the macroscopic and experimentally observable time \cite{bochner,carnap}.

In many physical and biological systems the variation of a signal is delayed by the presence of internal structures and processes that are not easily accessible to experimental observations or are neglected \cite{delvenne}; therefore, in these cases, we experimentally deal with signals showing extra layers of complexity. Retrieving the processes $X$ and $T$ does not represent an easy task especially for real or in silico 
 biological models. One of the limitations is the quantification of data available for statistical analysis. Another limit is that the sampling time, generally, does 
not coincide with the times $t^\star_i$; Third, experimentally it may be not 
 evident what $X$ represents and if any perturbation in the signal should be considered relevant or just noise. For these reasons, the disentangling of the subordinated process is not unique, and we preferred to address the problem numerically.  
Our approach requires to identify critical events in the system under analysis. 

In our ABM the number
of the molecules in specific states fluctuates due to random microscopical interactions between the molecules. These fluctuations are driven at meso/microscopical scale by interactions among specific single particles, but experimentally they are elusive and present many technical problems in terms of instrumental resolution or capability of tracking large number of single particles.
Rapid variations in the number of molecules in specific states affect the dynamic of the model, and their generation-relaxation towards a 
changing equilibrium condition is characteristic of the model itself. Hence, it is reasonable for this ABM to define critical events as the large and abrupt fluctuations in each molecular specie.
In order to detect the events for a molecule $i$, first, we filtered the time series, $y_i$, with a moving average over a time lapse $\tau_1$ on each state variable , such that $y^\prime_i(t)= \widehat{A}_{\tau_1}y_i(t)$, then we applied a second time the operator $\widehat{A}_{\tau_2}$ to the variation of each molecular specie $i$ from its estimated mean $\Delta y^\prime_i(t)=y^\prime_i(t)-y_i(t)$ the result of which is $y^{\prime\prime}_i(t)= \widehat{A}_{\tau_2}\Delta y^\prime_i(t)$, where the intervals of time are constrained by $\tau_1>\tau_2$. The first moving average has been used to de-trend the time series from the signal (intended as the component with slower dynamics). The subsequent second moving average has been applied to determine the amplitude $\sigma_i(t)$ of the second moment of the non-local noise fluctuations $\Delta y^\prime_i(t)$ central with respect to $y^{\prime\prime}_i(t)$. The term non-local is intended as non-local in time, and it includes all the times belonging to $[t-\tau_2/2,\ t+\tau_2/2]$ for each time $t$.

The time series $\Delta y^\prime_i(t)$ fluctuate around zero. The Hilbert transform, defined as the integral operator
$$
\widehat{\mathcal{H}}=PV\int_{-\infty}^\infty (\cdot)\ \frac{d\tau}{\pi(t-\tau)},
$$
has been used on the de-trended signals to derive the symmetric envelope of the de-trended signal, $y_{\mathrm{env}}(t)=\left|\widehat{\mathcal{H}}\ y^\prime_i(t) \right|$, from which it has been subsequently detracted. The result, $y^{\prime\prime\prime}(t)=y^\prime(t)-y_{\mathrm{env}}(t)$, is a positive defined time series with enhanced prominent peaks and dumped valleys close to the zero axis, \fref{fig_sub}. 
For each component $y^{\prime\prime\prime}_i$ of the signal, the peaks are detected as events $\mathcal{E}$, if the signal minus one standard deviation of the fluctuations drops after each peak, $p_n(t)$, below the corresponding preceding peak value:
$$
\mathcal{E}=\{p_n(t_1), \exists\ t_2\ |\ y^{\prime\prime\prime}_i(t_2)-\sigma_i(t_2)<y^{\prime\prime\prime}_i(t_1)\}.
$$

The component $y^{\prime\prime\prime}_i$ has regions of small values
; the beginning and ending of those regions are detected as events if the subtraction of one standard deviation of the fluctuations to the signal drops below or increases above the zero.
The WTDs
between peaks or the time extent of peaks are considered. To avoid artefacts related to fixed bin size and to be able to compare distributions obtained with different numerical parameters (in general having different binning sizes), we use a Kernel Density Estimator (KDE) defined as $\mathrm{KDE}(t)= \frac{1}{w}\int_t^{t+w} K(t-\tau) (\cdot) d\tau$ such that $\int_{-\infty}^\infty K(t)\ dt=1$, and $w$ is the support of the operator.



\

\subsection*{Signal post-processing}


The directing process, defined above as one component of the subordinated signal and here numerically obtained by detection of critical events, provides information on the recurrences of large fluctuations of the number of molecules rapidly reaching or departing from a given node of the network, \fref{fig_sub}. 
The waiting time distribution (WTD)
of the directing process depends on the chosen parameters; therefore, it is representative of the specific dynamics of the model. 


In our model, all the reaction occurrence times 
have a maximum
time limit, 
but not all interactions at the level of single agents are ruled by a single time distribution. Indeed, binding of molecules is driven by Brownian motions, while dissociation and relaxation of molecules are uniformly distributed, or have a constant value. Furthermore, the external perturbation maintains the system in a out of equilibrium condition. 

Consequently, the directing processes of many molecules largely depart from a classical Poisson process, where the critical events obtained by numerical detection would be gamma distributed, and characterize by multimodality. 

In our approach 
the integral distribution is first derived  by weighing each inter-time $\tau$ between two consecutive events, $y^{\prime\prime\prime}_1(t)$ and $y^{\prime\prime\prime}_2(t+\tau)$, with the amplitude of the fluctuation $y^{\prime\prime\prime}_1(t)$.  The area $\tau \times y^{\prime\prime\prime}_1(t)$ 
is 
a measure of the minimum cumulated time that molecules in a given state $i$, which have been involved in the same critical event, are going to spend in any state different from $i$. We can see it as the cumulated dispersion time of molecules sharing the same kinematics by departing from the same node of the network.  

The non normalized distribution gives the amount of the total number of molecules with the same averaged dispersion time found inside the entire duration of the sampled signal.
We would like to stress that all the signals have been de-trended and positively defined.
The  WTD, $\psi(\tau)$ is the probability density obtained by normalizing the integral distribution to 1.
In order to verify the repeatability of the results, for each set of parameters, we repeated 10 independent simulations. For each simulation, we computed the distributions, and we used the 10 independent results to derive the corresponding  punctual confidence interval, (notice that the top and bottom of confidence intervals around the WTDs are not probability functions).

Given the large amount of data simulated and results, we restricted the analyses prevalently on integrins, the RAF-MEK-ERK module of the molecular pathway and RUNX2. Our decision is also related to the importance of these proteins and their interactions in transduction of the mechanical forces from the membrane to the nucleus of the osteoblast as confirmed in the literature. \/*[].*/

\/*The functional form of the external perturbation is a periodic square wave which can be completely characterized by the amplitude and the period, while the phase is zero, and the minimal magnitude is set to 100 \fu. We have used periods which are shorter than the total duration of the simulation and a case in which the period is larger than the simulation, meaning the stimulation applied is constant. The external force has the role of activating the mechanotransduction cascade and the dynamic of the system which, otherwise, would remain trapped in its initial condition. For the period analysed, we have noticed the perturbation functions as a switch. When the perturbation goes at its minimum, the activation of the proteins also  goes to zero, and the other molecules in the pathway follow the same dynamics. For all the molecules which are downstream in comparison to the integrins, the larger their distance is, in terms of molecular networks, from the mechanical stimulus, the larger their delays are in following the dynamics of the external perturbation.*/
%
%
%
%
%






\subsection*{Simulations and sensitivity analysis}

The generic  ABM platform
FLAME (Flexible Large-scale Agent Modelling Environment) \cite{flameweb,bajo,paul}, 
capable of generating C based parallelisable code executable on different parallel hardware architectures, was used to create and simulate the model.
FLAME implements a discrete time fixed 
sweep update scheme meaning, in the simulation, the time advances at each iteration of a fixed amount defined as “time unit” in which all agents are updated only once in an internal defined order \cite{fate,gizhub}.
Our simulations start with \nagent agents evolving with a unit time step corresponding to 1 second for a period of $9\cdot10^4\,s$ (approximatively $24\,h$). 
The initial condition for the position and velocity of the agents has been chosen at random from a uniform distribution.
The mechanical load used to perturb the system is mimicked by a periodic square wave function defined by its magnitude $M$ and oscillation period $\period$. 

The proportions of agents belonging to specific molecular species, at time $t=0$, are constant among all the simulations performed, and their respective values are shown in \tref{proportions}. The values of the parameters for the baseline simulations are shown in \tref{base_parameters}. 
The simulations were run at the SHeffield Advanced Research Computer cluster (SHARC) on 
machines with 2 x Intel Xeon E5-2630 CPUs and 64GB of RAM.
Each simulation required 8GB of RAM, 96 hours of execution time and approximatively 500GB of hard disk space.   

The effect  on the fluctuations  of the molecular populations were investigated by changing the magnitude $M$ and oscillation period $\period$ of the external perturbation and the molecular complex dissociation  and relaxation times $T_{\bullet\ \rightarrow\ \bullet}$ for: 
%
$\RA{\s41 \rightarrow \s42}$,
$\RA{\s42 \rightarrow \s40}$,
$\RA{\s51 \rightarrow \s52}$,
$\RA{\s43 \rightarrow \s42}$,
$\RA{\s52 \rightarrow \s50}$,
$\RA{\s53 \rightarrow \s52}$
%
%
%
(\fref{fig_mechnet2}). 

%
%
%

All combinations for $N_M=2$ values of the perturbation's magnitude, $N_\period=3$ values of the perturbation's period, and $N_{\RA{\phantom{}}}=5$ values for each of the 6 dissociation/relaxation times has been analysed, and the spanned values of the parameters are reported in \tref{mut_parameters}.
In order to account for the repeatability of the ABMs, each simulation has been repeated 10 times for a total of 1800 simulations. Analyses and other results derived by the simulations are stored in FIGSHARE/Google\_Drive (link).

\subsubsection*{Phenomenological model and constraints}
In our biological molecular model, the number of simulated molecules inside an osteoblast is much smaller than in the real biological system. The reasons for such a decision is that running ABMs requires large resources in terms of memory and computational costs. This represents an upper limit in the number of agents which can be simulated. Another limit is given by spurious behaviours introduced by finite size and finite volume effects \cite{ascolani2,rhodes} representing an inferior boundary condition for the number of agents in the model. In our case, we have an average increasing number of agents during the simulation, which limited us in the initial maximum number of molecules. Nevertheless, the model does not explicitly  include any finite volume exclusion among molecules. Furthermore, the proportions of some molecules such as integrins in a specific state on the cell membrane are strongly driven by the external perturbation independently from the total amount of integrins simulated; this phenomenological aspect overrides many finite size effects. Consequently, we are in power to state that we can rescale the concentrations of the molecules and maintain the same averaged dynamic of the system by rescaling the interaction range \irange of the molecules accordingly \cite{pogson,rhodes}. 




\section*{Results and discussion}
\subsection*{Waiting time distributions and spectrograms}
\subsubsection*{Dissociation times}
During interaction between  MEK and RAF, MEK is phosphorylated and consequently activated; after dissociation of the compound, activated MEK propagates the mechanotransduction signal. In \fref{fig_RADP10edited}, we found the fluctuations of the complex MEK-RAF presents two distinct dynamics depending on the dissociation time. For dissociation time around $\textrm{\RADP10}=10\, s$, the WTD  is a bimodal distribution with modes around the recurrence times $\tau_1=4\, s$ and $\tau_2=9\, s$, \fref{fig_RADP10editedC}. The distribution $\psi(\tau_1)$ is approximately 15\%  larger than in $\tau_2$. 
When \RADP10 is increased from $90\,s$ to $1320\,s$, the dynamics of the fluctuations changes completely; the peak at $\tau_1$ becomes negligible and  the distribution has only one relevant maximum at $\tau=9\, s$. At  $\textrm{\RADP10}=1320\,s$ the integrated distributions and the WTD show that  a large portion of the intertimes' fluctuations are close to the mode and the occurrence of critical events is more regular.
For $\textrm{\RADP10}=10\,s$, we also see the number of molecules with same dispersion time is below $200$ for any delays 
and any  period of the mechanical stimulation, while it is larger than $400$ at $\tau_2$ for dissociation times $\textrm{\RADP10}>10\,s$.     

From the analyses of the occurrence of critical events, 
all mRNA sequences bound to ribosomes show similar  dynamics.
As illustrated in \fref{fig_ageing_RADP13_P1000} and  \fref{fig_ageing_RADP1113_P200000},  for $\textrm{\RADP13}=1320\,s$, the WTDs of mRNAs bound to ribosomes are trimodal and the corresponding peaks are at $\tau_1=4\,s$, $\tau_2=10\,s$ and $\tau_3=18\,s$, respectively.
The shapes of the WTDs remain constant at all the epochs simulated with the exception of the mode at larger recurrence time $\tau_3$ which tends to fluctuate due to stochastic noise. 
These properties hold true for P equal to $1000\,s$ and $5000\,s$ (see \fref{fig_ageing_RADP13_P1000A} and \fref{fig_ageing_RADP13_P1000B}), but for constant mechanical loads, the WTDs loose their trimodality and become definitely bimodal after the age $t=5000\, s$ as shown in \fref{fig_ageing_RADP1113_P200000C} and \fref{fig_ageing_RADP1113_P200000D}.
Similarly, in \fref{fig_ageing_RADP1113_P200000A} the WTDs for $\textrm{\RADP11}=1320\,s$ and $P=200000\, s$ have three modes at $\tau_1=4\,s$, $\tau_2=10\,s $ and $\tau_3=18\, s $. After the system reaches the age of $t=5000\, s$, the mode at $\tau_3\/*=18\, s*/ $ disappears while the other two modes remain at equal $\tau$,  \fref{fig_ageing_RADP1113_P200000B}. In all other cases, the WTDs of mRNAs bound to ribosomes are bimodal. 
Systems under constant mechanical loads with large activation times of MEK or ERK have a dynamics characterized by 
bursts of translation occurring at intertimes smaller than 10 seconds. On the contrary, for periodic mechanical loads, there is a larger probability that abrupt variations in mRNA translation are separated by intertimes twice as long.    

Free mRNAs have completely different WTDs compared to their bounded counterparts. Generally, all the free mRNA WTDs have modes at intertimes equal to $\tau_1=5\,s$, $\tau_2=41\,s$ and $\tau_3=55\,s$ which are the same at all epochs for any of the dissociation/relaxation times considered and all values of $P$, $M$ and dissociation/relaxation times $T_{\bullet\ \rightarrow\ \bullet}$ simulated, \fref{fig_free_mRNA}. 
The intertime range between $\tau_1\/*=5\,s*/ $ and $\tau_2\/*=41\,s*/ $ is characterized by multiple peaks changing with the age of the system, \fref{fig_free_mRNAA}.
Except for $\textrm{\RADP13} < 300\, s$, in the WTDs of free mRNAs there is a stable peak around $\tau_4=10\,s$ and an oscillating tail which slowly decays till the next stable mode at $\tau_3\/*=41\,s*/ $. Instead, for values of \RADP13 smaller than $300\,s$, there is no stable peak at $\tau=10\,s$ and decaying tail becomes a long fluctuating plateau, see \fref{fig_free_mRNAB}. 
Consequently, faster activation of ERK induces a more focused transcription of mRNAs, while at larger values of \RADP13, there is a wide set of intertimes between $\tau_1\/*=5\,s*/ $ and $\tau_2\/*=41\,s*/ $ with high probability which causes nonrhythmic rapid variations of mRNA production.      
It is important to emphasis that  constancy of behaviours among all types of mRNA, which, in the ABM, are 
identically interacting molecules 
\/*molecules interacting in the same way*/ and thus are dynamically indistinguishable, shows the consistency of the fluctuation analyses and the robustness of the method.

Reducing the values of $T_{\bullet\ \rightarrow\ \bullet}$ causes  an increase of molecular interactions and a corresponding effect on the number of fluctuations in the signals. Such phenomenon can be seen in non normalized histogram of bounded molecular complexes like RAF+MEK, \fref{fig_RADP14t1000editedA} and \fref{fig_RADP14t200000editedA}.  Instead, free molecules show a direct relation between the size of the fluctuations and the dissociation/relaxation value, \fref{fig_RADP14t1000editedB}-\subref{fig_RADP14t1000editedC} and \fref{fig_RADP14t200000editedB}-\subref{fig_RADP14t200000editedC}.
Nevertheless, we have observed that the tendency of higher amount of events for smaller value of $T_{\bullet\ \rightarrow\ \bullet}$ does not always hold, see \fref{fig_saturationA}-\subref{fig_saturationC}. In such cases, after the initial increase of the number of events,  there is saturation. Decreasing  $T_{\bullet\ \rightarrow\ \bullet}$ further,  there is a variation in the dynamics, and the quantity of critical events becomes directly proportional to the value of the relaxation/dissociation time.

\subsubsection*{Mechanical perturbations}


The functional form of the external perturbation is a periodic square wave which is completely characterized by the amplitude $M$ and the period $\period$, while the phase is set to zero, and the minimal magnitude is equal to $100\,\mu Pa$, see \tref{base_parameters}. 
In the range of values of $\period$ investigated, two were much shorter than the total duration of each epoch, and in one case
the period was set such that the stimulation applied was constant for the whole simulation. The external force has the role of activating the mechanotransduction cascade and the dynamic of the system which, otherwise, would remain trapped in its initial condition. 

For all the analysed epochs and the simulated periods, the results indicated that the perturbation functions as a switch. When the perturbation goes to its minimum, 
integrin activation  approaches zero, and the other molecules in the pathway follow the same dynamics. For all the molecules which are downstream in comparison to the integrins, the larger their network distance is, the larger their delays are in following the dynamics of the external perturbation (\fref{fig_mechnet1}).


The passage of the mechanical loads from high to low and vice versa is of the order of 1 time unit (1s), and the duration $\period$ of each regime is at least $1000\,s$ long (see \tref{base_parameters} and \tref{mut_parameters}). 
Indeed, all the WTDs analysed have a support shorter than all $\period$ considered, and in 
\fref{fig_RADP10edited}%
\,-\,\ref{fig_ageing_RADP1113_P200000}%
, \ref{fig_RADP14t1000edited}%
\,-\,\ref{fig_ageing2}, the recurrence of events does not require a time larger than $100\,s$.  


The paucity and sparsity of the events characterized by the high/low transitions of the external perturbation do have a minor effect on the intertimes $\tau$ between molecular events even in cases where the dynamics of large quantities of molecules synchronize with the external signal. On the contrary, during the time period $\period$, molecules have the chance to form compounds or dissociate multiple times, and these events, induced by the faster molecular dynamics, account for most of the statistics shown in the WTDs.

The increase of the magnitude $M$ of the perturbation resulted in an increase of the number of molecules activated which can be clearly observed from the variations in the non normalized histograms, but this does not always correspond to a significant change in the distributions of the intertimes, see third row in \fref{fig_noise}.  

 
For active integrins,
increasing $P$ of a factor 40 resulted in reduction of the variability among independent repetitions of the simulations. In the first and second rows of \fref{fig_noise}, the WTDs with $P=5000\,s$ 
present a tail with a large confidence interval, 
and at large $\tau$, the shapes  largely fluctuate among different epochs. Instead, for $P=200000\,s$, the waiting times    
produce clear bimodal distributions with reduced confidence interval at large $\tau$ resulting in no tail and a stable shape during different ages. The reduction of noise on active integrin intertimes is strongly sensitive to the length of the period $P$, but not on the magnitude of the perturbation, see third row in \fref{fig_noise}. 
We noticed the sensitivity on $P$ of active integrin intertimes does not depend on the value of \RADP10, \RADP11 and \RADP12. Instead, when the ERK cycle is slowed down, corresponding to \RADP13, \RADP14 and \RADP15 larger than the baseline values, the WTDs' noisy tails disappear, and the distributions are the same for any value of the perturbation period.

Increasing the period $\period$ implies larger lapse of time in which 
the mechanotransduction is switched on. Consequently, a larger number of events is expected. If the dynamics of the system does not change sensibly during the oscillations of the perturbation, then the area under the histograms increases with the increase of $\period$, while the WTDs do not change.


\subsubsection*{Ageing}
The integral distribution and the WTD are quantities computed over a time period, named epoch, which are much larger than the support of the represented distributions and smaller than the total length of the simulation. Because the system is in a non equilibrium condition, we cannot state that the WTD does not depend on the epoch; therefore,  the WTD can be rewritten as $\psi_t(\tau)$ where $t$ is the initial time of the epoch of fixed extension used for the computation.  Larger epochs reduce the lack of statistics and noise, while smaller epochs allow us to compute $\psi_t(\tau)$  by choosing the age $t$ from a larger range of available values. We tested different epoch sizes, and we compromised for a period of 10000 seconds.   

Ageing is characteristic of all the interacting agents, even though it is not so easy to estimate the effects of the passage of time in the histograms and WTDs, due to large confidence intervals and noise therein. 
Indeed, for some parameters and molecules, both the histograms and WTDs show alternate crescendo/decrescendo patterns and
irregular cyclical variations. Even if the system is maintained in an out of equilibrium condition, 
the dynamics of the system responds and adapts accordingly to a mechanical perturbation.


Ageing is a direct consequence of the initial conditions; nevertheless, 
given that the system is perennially far from the equilibrium, ageing is prevalently due to a variation of regimes in the dynamics of the system. Therefore, the state of the system at $t=0$ should not be seen as a static initial condition, but as if the system reached such a state due to a specific long standing dynamics. 
Indeed, we have simulated a case where the cell had been stimulated with two consecutive trains of oscillating perturbations with different frequencies. The resulting histograms show there was no ageing 
at the end of the first perturbation and that ageing appeared during the second regime. 





Among all molecules analysed, transcribed factors are characterized by ageing with a most evident and clear progression. 
From the  integral distributions in \fref{fig_ageing2},
we observe that, with the increasing of age $t$ there is a reduction on the number of abrupt fluctuations. A similar ageing effect appears in all the mRNAs bound to ribosomes. On the other hand, in many of the cases analysed, the corresponding WTD does not show sufficient discrepancies between different epochs. 
This means all the dispersion times decrease of the same amount as time passes, and the ageing effect is reduced to a slowing down of the system dynamics.
As a direct consequence of the fixed amount of total ribosomes simulated in the cytoplasm, ribosomes in their free state also present an ageing effect, but in this case, the dispersion times increase with the age of the system. 
For free mRNA, the variations with the age $t$ of the dispersion times change at different rates depending on the intertime value $\tau$; consequently, their WTD does not preserve the shape, but changes from a trimodal  distribution to  a bimodal distribution as time passes, \fref{fig_ageing_RADP1113_P200000}.  

\section*{Conclusions}

The dynamics of various molecules involved in the  bone cell mechanotransduction has been addressed in the framework of ABM.
Preliminary analyses on the resulting signals \/*have shown*/ showed the external perturbation functions as a switch, but did not suggest any further synchronization on time scales smaller than the time period of the external mechanical loads. The amount of molecules in time were driven by a slower dynamics frequently interrupted by asynchronous large fluctuations.   
Such behaviour is a common characteristic of subordinated processes. The lack of synchronization has suggested retrieving the directing processes of \/*such*/ the signals and analyse their distributions in function of different parameters.


The method of disentangling the principal process and the directing process is not unique, and it depends on the modality used to detect the \/*occurrence of*/ events. 
\/*Furthermore, the developed method*/ The approach developed in this work permits dealing with systems in out of equilibrium conditions resulting from a particular initial state or due to the external perturbations, and it could be applied to different types of time series.
This \/*approach*/ method has been used to study the dynamics of osteoblasts' mechanotransduction which connects the extra cellular properties with the intracellular signals and includes a large set of diffusive chemical reactions. 
By identifying the occurrence of an event with multiple reactions at the molecular scale happening at the same time, the proposed numerical technique allowed us to derive the directing process for each molecular class.
Differently from more traditional \/*other*/ approaches, where noise \/*in the signals*/ has been \/*relegated to*/ \/*considered*/ adopted as a \/*a possible*/ measure to quantify the error around the expected values of time dependent signals,  here, the \/*large and*/ abrupt fluctuations departing from the trend have been explored and analysed in order to extract hidden information characteristic to each class of molecule and to the process dynamics.

The WTD of many classes of molecules presented multimodality which highlight the presence of preferred intertimes in the occurrence of events. \/*The variation of*/ Changes in the relaxation or dissociation times between molecules in the RAF-ERK-MEK module  showed variations in \/*the*/ WTDs of molecules belonging both to the classes near one of the perturbed edges and to other modules of the mechanotransduction network.
For example, under  a $1000\, s$ time period perturbation, the WTD of RAF bound to MEK was characterized by a bimodal distribution for low values of active MEK dissociation times, while it became unimodal at larger value of the same parameter. 
Many of the mRNAs' WTDs were affected by ageing, \/*resulting in*/  a variation of the shape from a bimodal to trimodal distribution, showing the emergence of a new preferred intertime  of occurrence of event.    
Changes in the WTDs' shape  are proved to be significant by local error estimation of the distribution.

Variation in the modes implies variations in the dispersion times between chained events of the molecular network. 
Hence, the WTD of a specific molecule is a dynamics' marker which can be used as a signature for pathological conditions \/*signature*/ helping to design drugs
\/*capable of maximize their target interaction, 
in order to maximize the interaction with the target antagonize the chained reactions and be target selective selective.*/
capable of improving target selectivity, maximizing the interaction with targets and antagonizing chained reactions.
Our method of analysis is a suitable tool to study in vitro experiments tailored to investigate the effects of drugs with different properties like mass, diffusion coefficients or hydrosolubility.
Thanks to FLAME's parallelization and the genericity of the developed method, \/*our work on mechanotransduction in osteoblasts,*/ aspects like the Parathyroid hormone or inflammatory factors involved in bone remodelling and disease \/*bone mass*/  could be easily explored.  
%
Furthermore, the analyses done could be easily extended \/*is method would also be useful*/ to identify the principal processes of the molecules which describe the dynamics of the molecular baseline occurring at discrete time units. Such dynamics identify discrete variations which can be linked to overall survival and worsening/improved survival analysis \/*of characteristic events in the evolution*/ of diseases.



\begin{backmatter}

\section*{Competing interests}
  The authors declare that they have no competing interests.


\section*{Acknowledgements}
 This study was funded by the Engineering and Physical Sciences Research Council (EPSRC, Frontier Multisim Grant: EP/K03877X/1) 

\bibliographystyle{bmc-mathphys} 
\bibliography{bmc_article}      




\definecolor{greeno}{HTML}{007826}
\definecolor{greenoo}{HTML}{66FF99}
\definecolor{pink}{HTML}{E2D2E2}
\definecolor{yelloww}{HTML}{CC9900}
\definecolor{yellowww}{HTML}{FFFF99}
\definecolor{orang}{HTML}{FE7920}

\section*{Figures}

\begin{figure}[h!]
	\centering
	\captionsetup[subfloat]{position=top,captionskip=-16pt}
	\subfloat[][]{\label{fig_mechnet1}
		\centering	
		\includegraphics[scale=0.4]{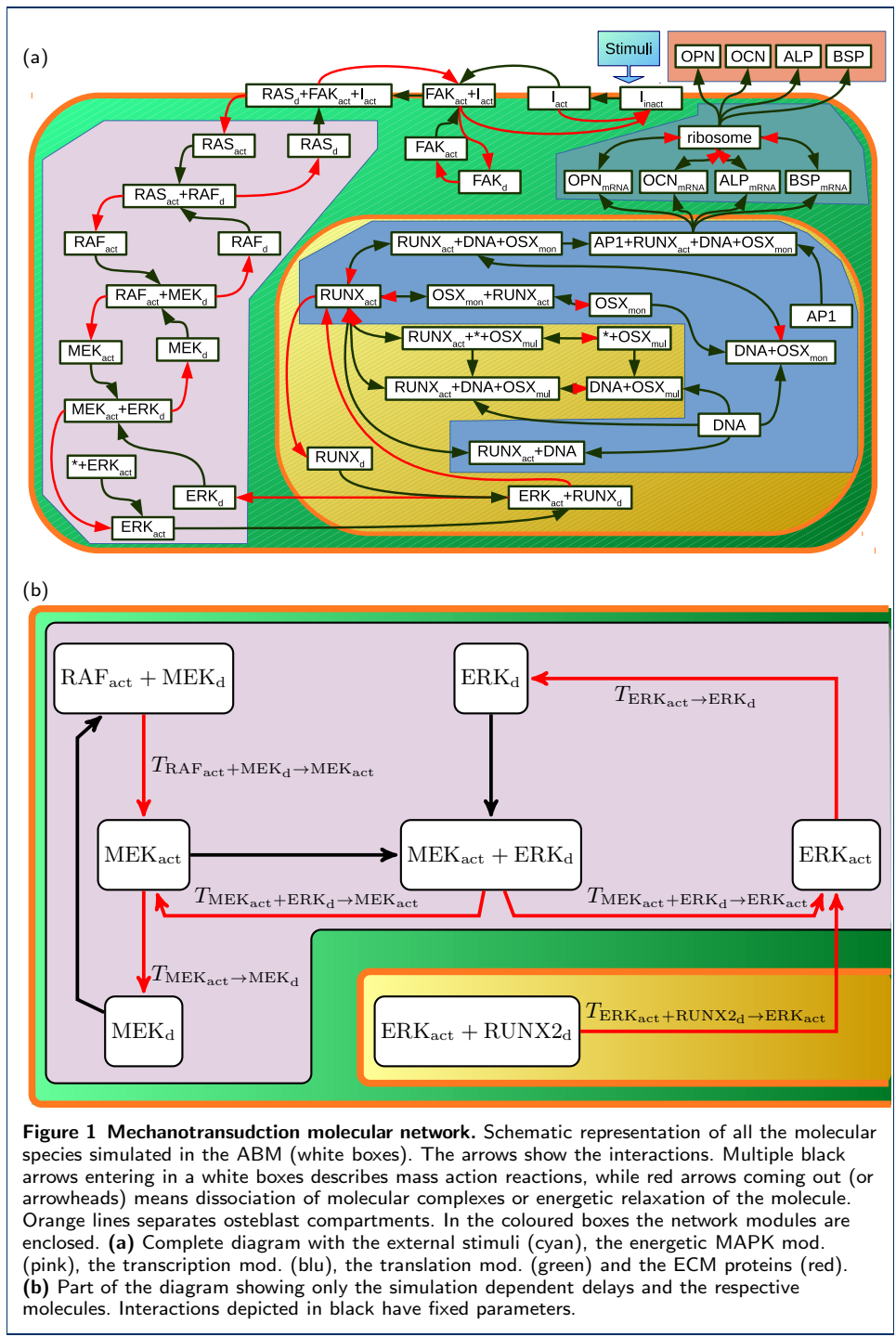} 
	}

	\captionsetup[subfloat]{position=top,captionskip=1pt}
	\subfloat[][]{\label{fig_mechnet2}
		\centering
	\begin{tikzpicture}[->,>=stealth',shorten >=1pt,auto,node distance=1.5cm and 3.cm,
	thick,state/.style={rectangle,fill=white, rounded corners, minimum width=1cm, minimum height=1cm,,align=center,text centered,draw,font=\sffamily\small\bfseries}]

	\shade[-,color=orang!0,top color=greenoo, left color=greenoo, bottom color=greeno, right color=greeno,opacity=1.] (-1.6,1) -- (10.7,1) -- (10.7,-6.1) -- (-1.6,-6.1) -- cycle;
	\draw[-,rounded corners,line width=0.1cm,color=orang,opacity=1.] (10.7,-6.1) -- (-1.6,-6.1)--(-1.6,1) -- (10.7,1) ;
	
	\shade[-,fill=pink,rounded corners] (-1.4,0.8) -- (10.7,0.8) -- (10.7,-3.6) -- (2.4,-3.6) -- (2.4,-5.8) -- (-1.4,-5.8) -- cycle;
	\draw[-,fill=pink,rounded corners] (10.7,-3.6)--(2.4,-3.6) -- (2.4,-5.8) -- (-1.4,-5.8)--(-1.4,0.8) -- (10.7,0.8)  ;
	
	\shade[-,rounded corners,color=orang!0,left color=yellowww,bottom color=yelloww, right color=yelloww,opacity=1.] (3.1,-4.2) -- (3.1,-5.8) -- (10.7,-5.8) -- (10.7,-4.2) -- cycle;
	\draw[-,rounded corners,line width=0.1cm,color=orang,opacity=1.] (10.7,-4.2)--(3.1,-4.2) -- (3.1,-5.8) -- (10.7,-5.8)  ;
	
	\node[state] (41)  {$\s41$
	};
	\node[state] (42) [below=of 41] {$\s42$
	};
	\node[state] (43) [right=of 42] {$\s43$
	};
	\node[state] (40) [below=of 42] {$\s40$
	};
	
	\node[state] (50) [above=of 43] {$\s50$
	};
	\node[state] (52) [right=of 43] {$\s52$
	};
	\node[state] (53) [below=of 43,xshift=-0.2cm] {$\s53$
	};

	\path[every node/.style={font=\sffamily\small},draw, ->,red,text=black]
	(41) 	edge[line width=0.05cm] node [left,anchor=west] {$\RA{\s41 \rightarrow \s42}$} (42)
	(42) 	edge[line width=0.05cm] node [left,anchor=north west,yshift=-0.2cm] {$\RA{\s42 \rightarrow \s40}$} (40)
	(43) 	[line width=0.05cm]-- (5.3,-3.4)--node [left,anchor=south, xshift=0.5cm] {$\RA{\s51 \rightarrow \s52}$}(9.6,-3.4) --  (52)
	;
	
	\path[every node/.style={font=\sffamily\small},draw, ->,red,text=black]
	(43) [line width=0.05cm]	-- (4.8,-3.4)--node [left,anchor=south,xshift=-0.2cm] {$\RA{\s43 \rightarrow \s42}$}(0.3,-3.4) --  (42)
	;
	
	\path[every node/.style={font=\sffamily\small},draw, ->,black,text=black]
	(50) 	edge[line width=0.05cm] node [left,anchor=north west] {} (43)
	(42) 	edge[line width=0.05cm] node [left,anchor=north] {} (43)
	;
	\path[every node/.style={font=\sffamily\small},draw, ->,black,text=black]
	(40) [line width=0.05cm] -- (-0.95,-4.6) -- (-0.95,-0.85)-- node [left,anchor=north] {} (41)
	;
	
	\path[every node/.style={font=\sffamily\small},draw, ->,red,text=black]
	(52) [line width=0.05cm] |-	     node [left,anchor=north east,xshift=-1cm] {$\RA{\s52 \rightarrow \s50}$}  (50);
	\path[every node/.style={font=\sffamily\small},draw, ->,red,text=black]
	(53) [line width=0.05cm] -|	     node [left,anchor=south east] {$\RA{\s53 \rightarrow \s52}$}  (52)
	;		
	\end{tikzpicture}
	}
	\caption{\csentence{Mechanotransudction molecular network.} 
		Schematic representation of all the molecular species simulated in the ABM (white boxes). The arrows show the interactions. Multiple black arrows entering in a white boxes describes mass action reactions, while red arrows coming out (or arrowheads) means dissociation of molecular complexes or energetic relaxation of the molecule. Orange lines separates osteblast compartments. In the coloured boxes the network modules are enclosed. \textbf{\protect\subref{fig_mechnet1}} Complete diagram with the external stimuli (cyan), the energetic MAPK mod. (pink), the transcription mod. (blu), the translation mod. (green) and the ECM proteins (red).  \textbf{\protect\subref{fig_mechnet2}} Part of the diagram showing only the simulation dependent 
		delays and the respective molecules. Interactions depicted in black have fixed parameters.
	} 
	\label{fig_mechnet}
\end{figure}

\begin{figure}
	\includegraphics[scale=0.6,angle=-90,trim={15ex 0ex 12ex 0ex},clip]{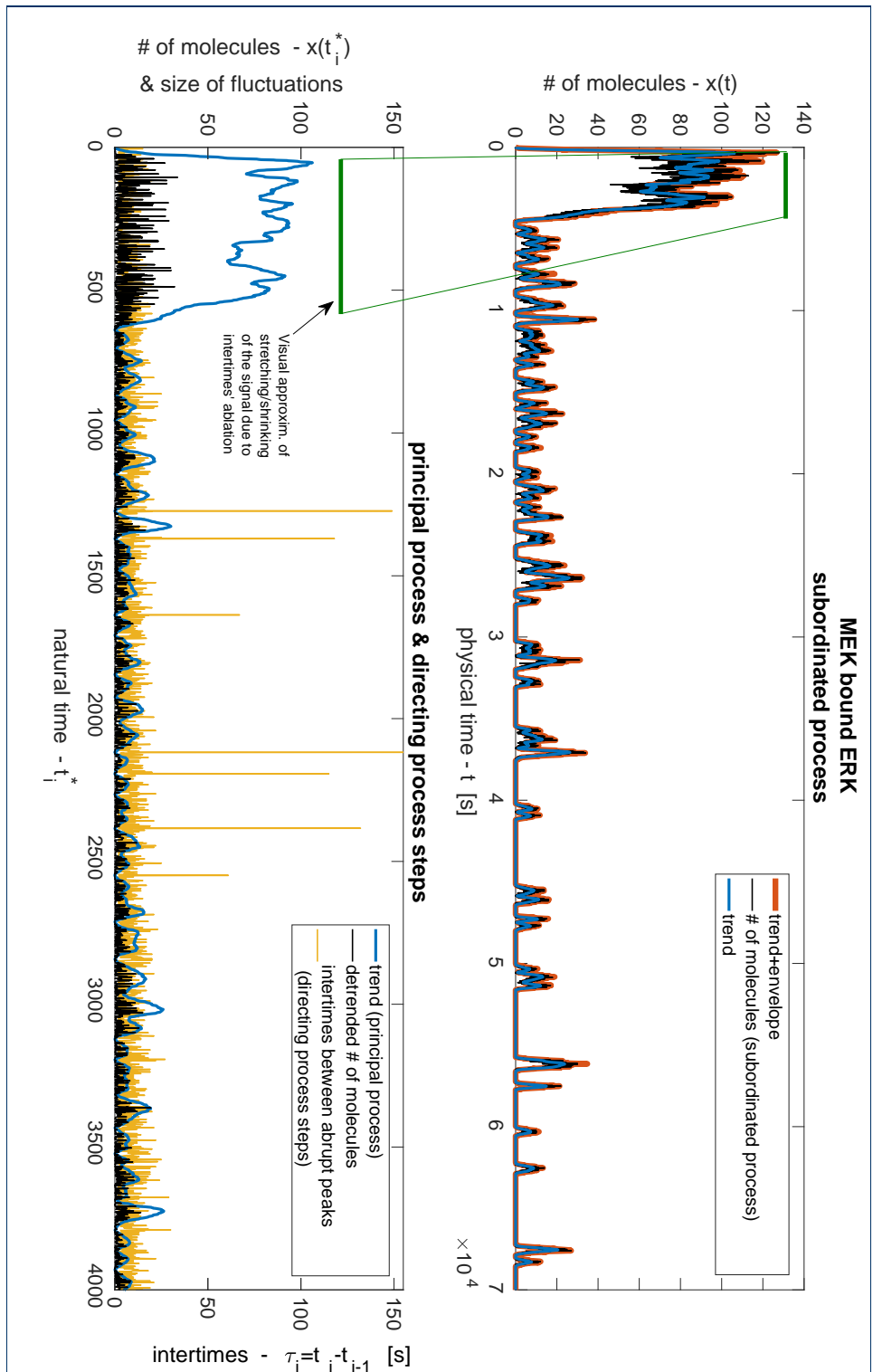}
	\caption{\csentence{Subordinated process.}
		On the top graph, the amount of molecules for the complex MEK+ERK in function of the time $t$ is shown . This curve represents the subordinated signal $x(t)$. The trend of the signal, derived by time average, and the envelope, derived by Hilbert transform, are used to detect the critical events. In the bottom graph, the intertimes $\tau_i$ between two consecutive critical events are shown in function of the physical time $t_i^\star=i$. The curve $t(t^\star_i)$, given by the sum of all the intertimes $\tau$ up to the i-th event, is a realization of the directing process. In the same graph, the curve  $x(t^\star_i)$, given by the sum of all the steps $\Delta x$  up to the i-th event, is a realization of the principal process. For convenience, in the bottom graph, the fluctuations around the trend are shown. In the curve $x(t^\star_i)$,  regions with high number of events resemble a stretched form of the subordinated signal while regions with few events resemble shrunk portions of the subordinated signal.     
	}
	\label{fig_sub}
\end{figure}

\begin{figure}[h!]
	\centering
	\captionsetup[subfloat]{position=top,captionskip=-18pt}
	\subfloat[][]{\label{fig_RADP10editedA}
		\centering
		\includegraphics[trim={-18ex 10ex 100ex 0},scale=0.265]{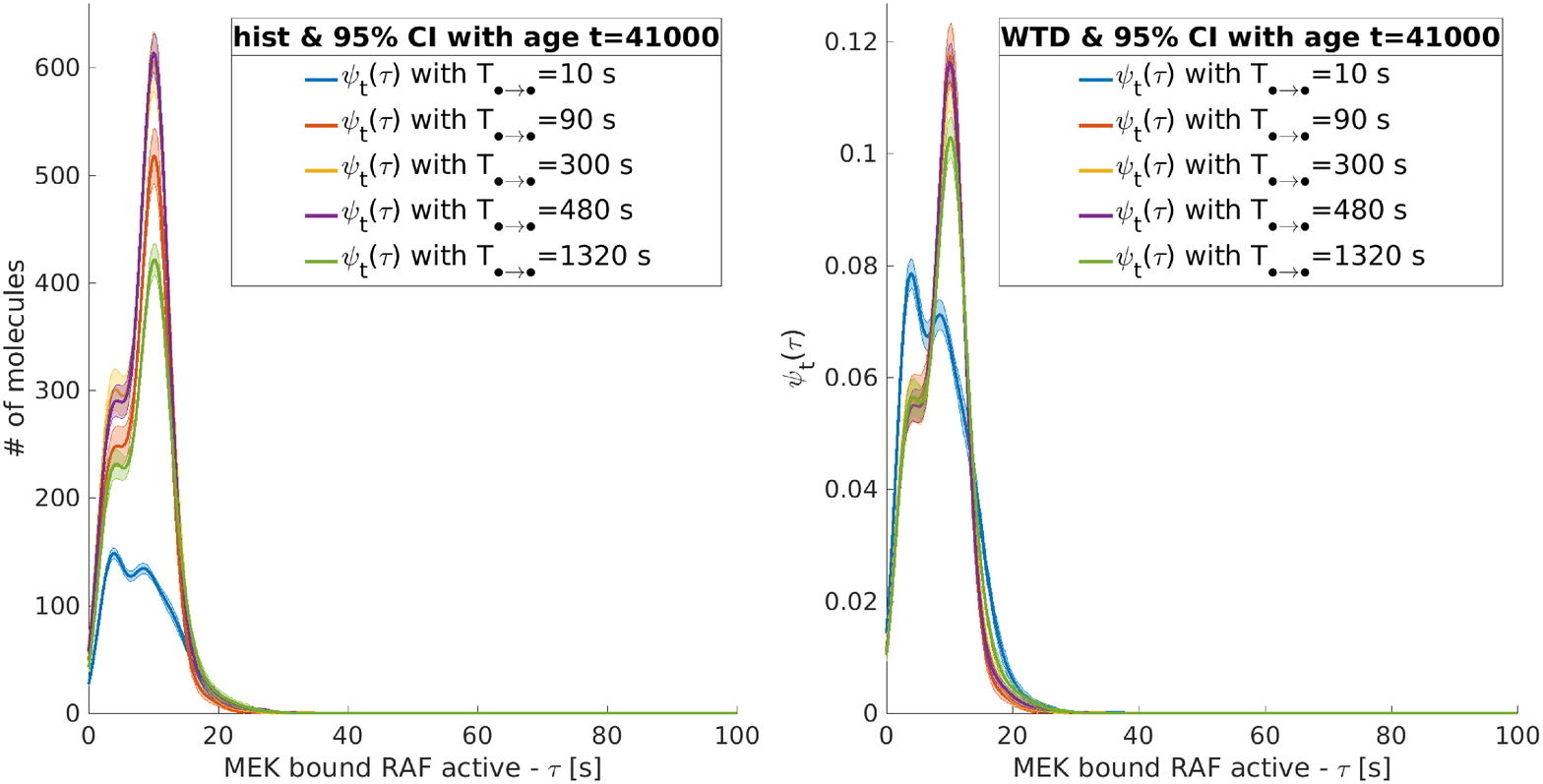}
	}
	\vspace{-4pt}
	\captionsetup[subfloat]{position=top,captionskip=-18pt}
	\subfloat[][]{\label{fig_RADP10editedB}
		\centering
		\includegraphics[trim={-18ex 10ex 100ex 0},scale=0.265]{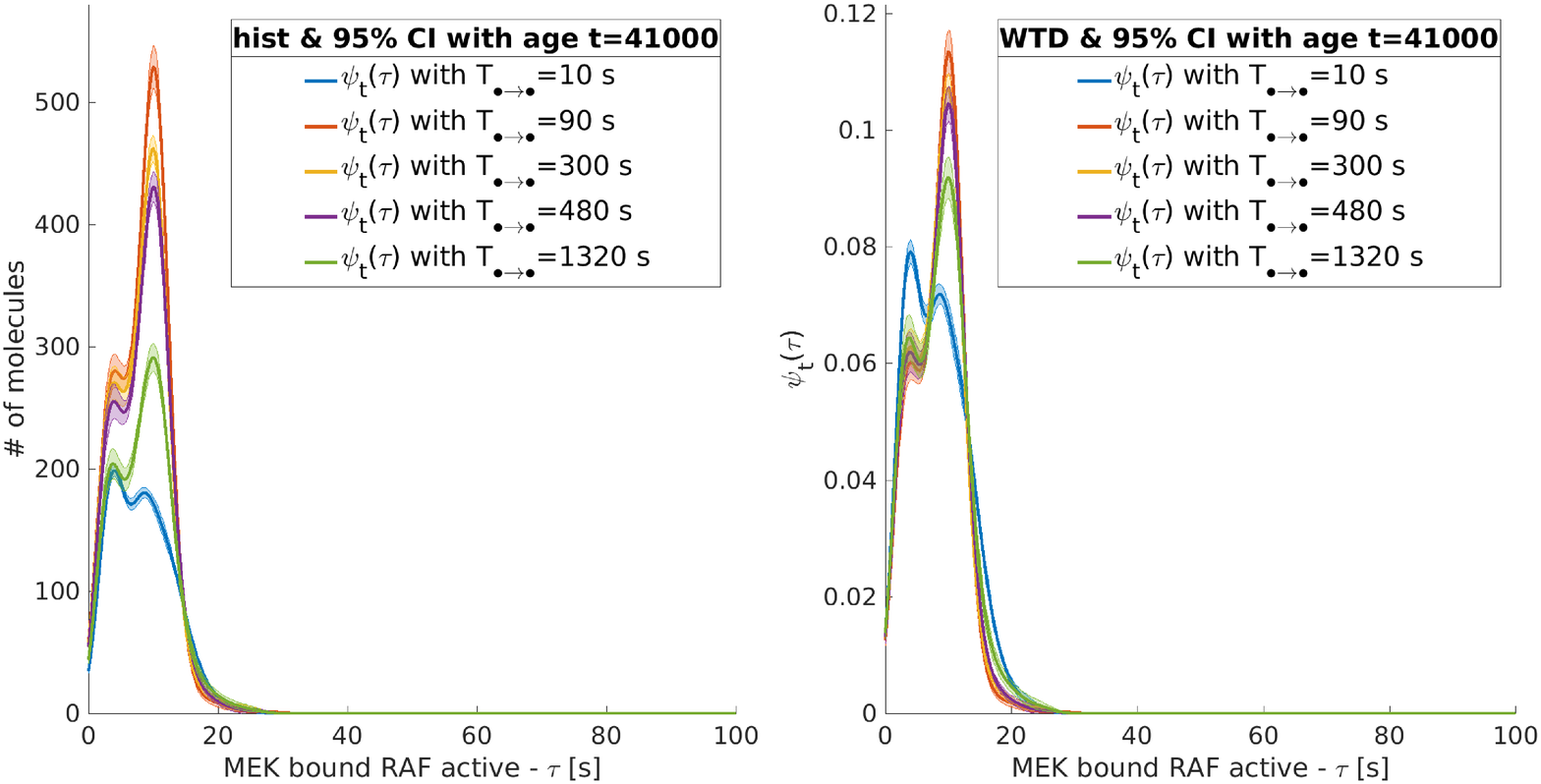}
	}
	\vspace{-4pt}
	\captionsetup[subfloat]{position=top,captionskip=-18pt}
	\subfloat[][]{\label{fig_RADP10editedC}
		\centering
		\includegraphics[trim={-18ex 10ex 100ex 0},scale=0.265]{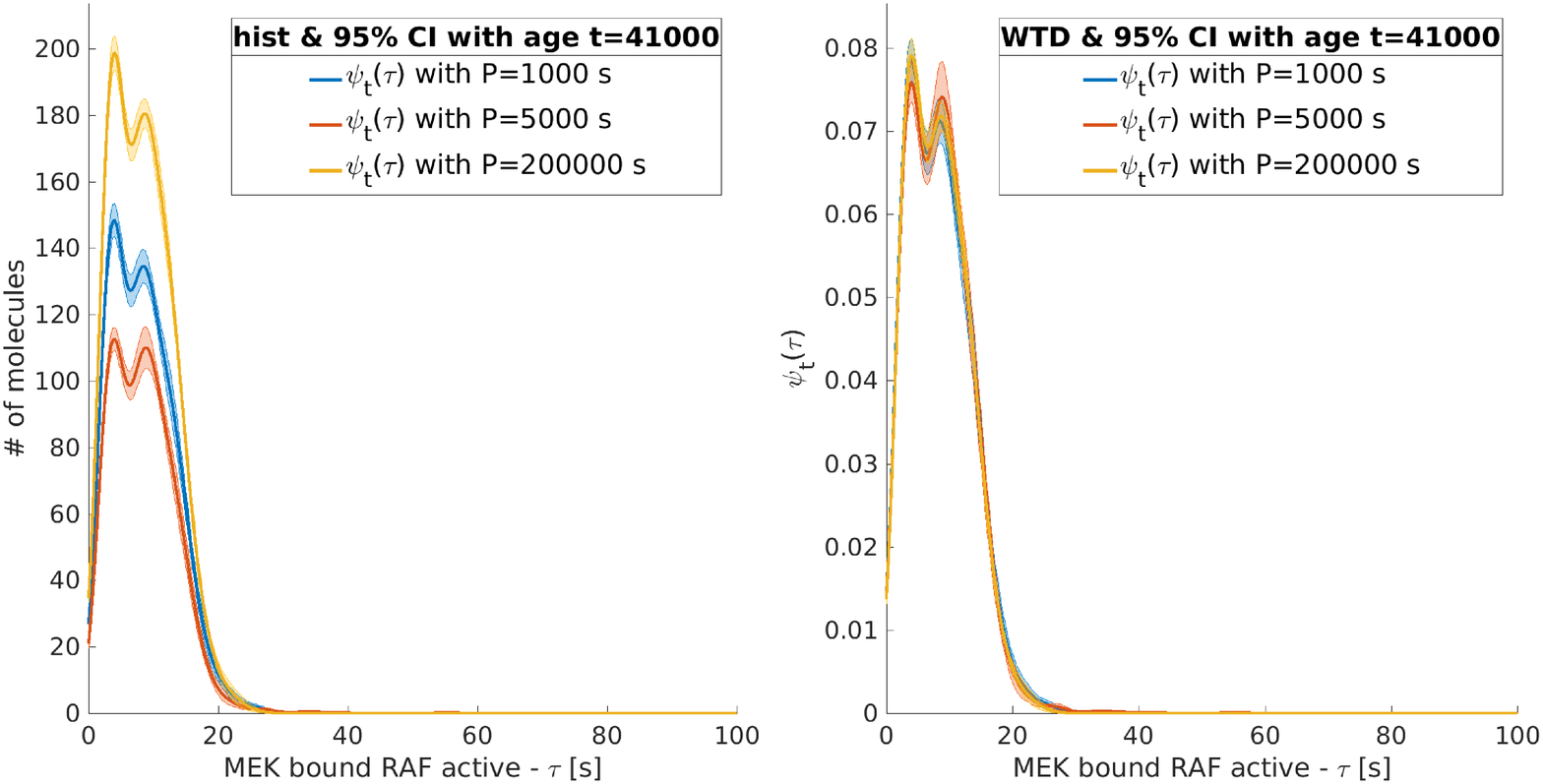}
	}
	\vspace{-4pt}
	\captionsetup[subfloat]{position=top,captionskip=-18pt}
	\subfloat[][]{\label{fig_RADP10editedD}
		\centering
		\includegraphics[trim={-18ex 10ex 100ex 0},scale=0.265]{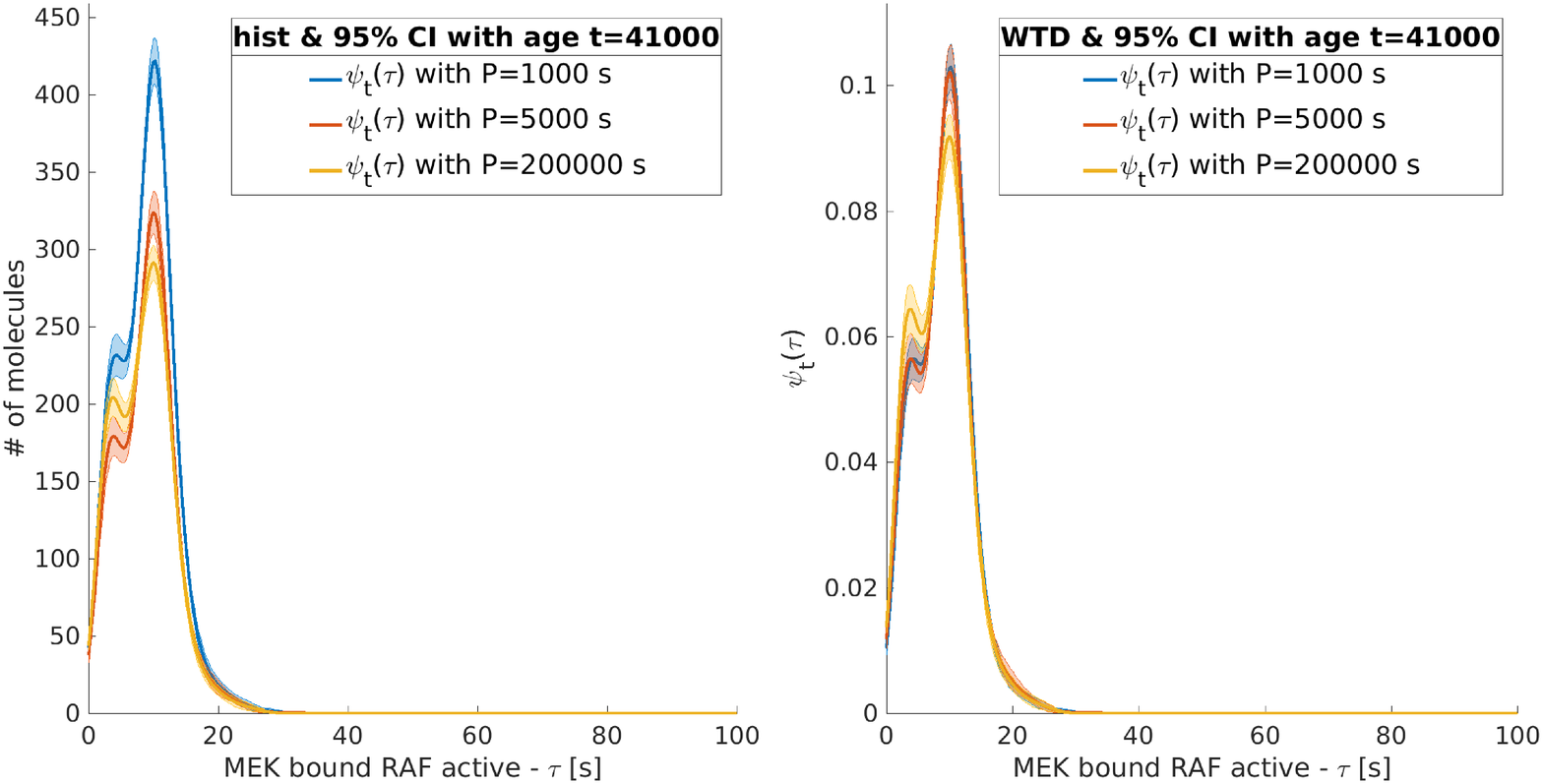}
	}
	\caption{\csentence{MEK bound RAF.} 
		Independently from the frequencies of the tested perturbations, the variation of \RADP10 leaves the MAPK sub module of the network unchanged exception for the MEK interacting with RAF which is highly sensitive to the parameter. Indeed, the distribution of recurrence of events for active MEK bounded to RAF is bimodal in $\tau=\{4, 9\}\,s$. For small value of $\textrm{\RADP10}\sim10\,s$ the distribution is mainly centred around $\tau=4\,s$ and it represents the main mode, while for larger value of \RADP10 the distribution's main mode is at $\tau=9\,s$. \textbf{\protect\subref{fig_RADP10editedA}} $\T=1000\,s$; \textbf{\protect\subref{fig_RADP10editedB}} $\T=200000\,s$; \textbf{\protect\subref{fig_RADP10editedC}} $\textrm{\RADP10}=10\,s$; and \textbf{\protect\subref{fig_RADP10editedD}} $\textrm{\RADP10}=1320\,s$.
	}
	\label{fig_RADP10edited}
\end{figure}

\begin{figure}[h!]
	\centering
	\captionsetup[subfloat]{position=top,captionskip=-30pt}
	\subfloat[][]{\label{fig_ageing_RADP13_P1000A}
		\centering
		\includegraphics[trim={-18ex 10ex 0 10ex},scale=0.265,clip]{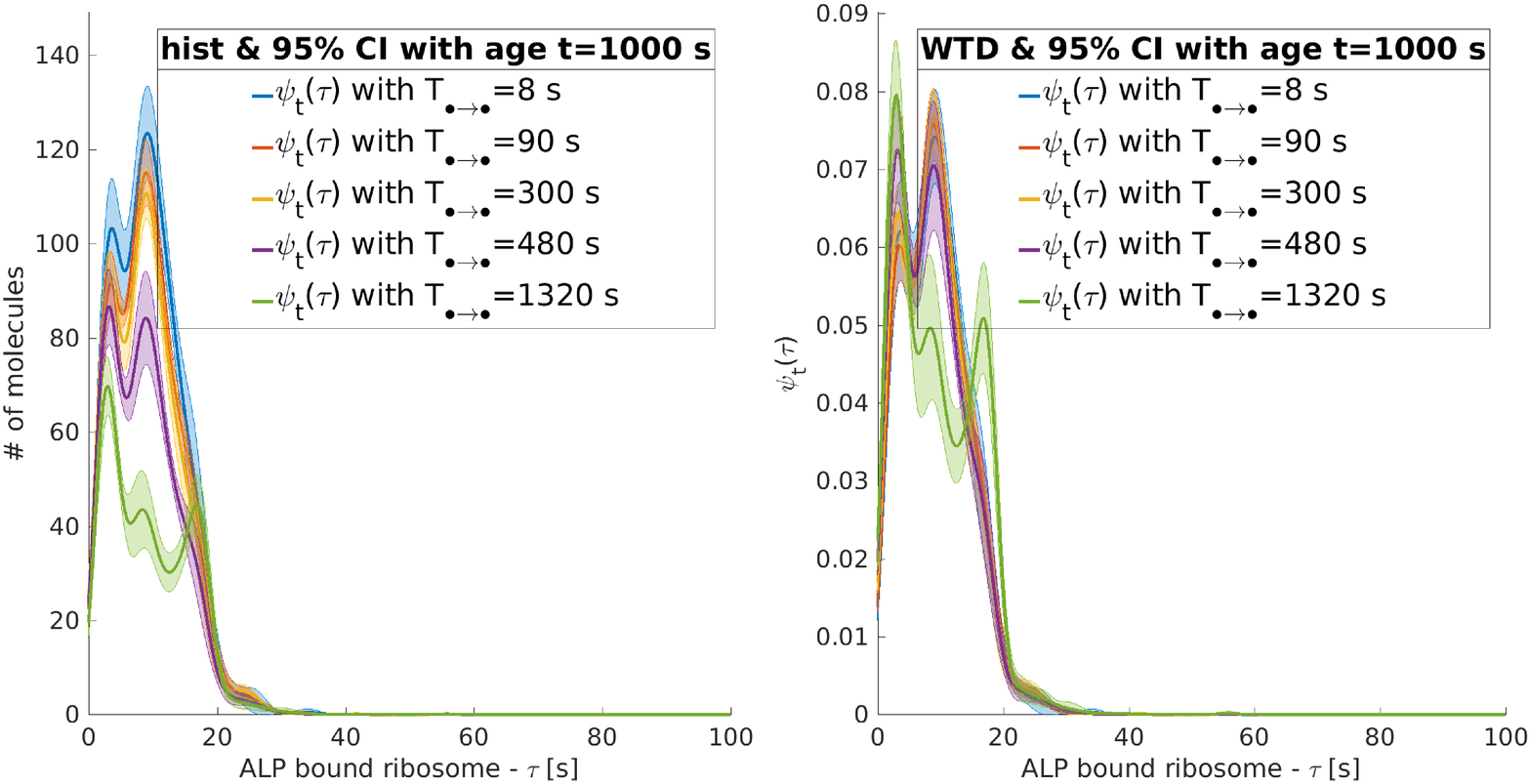}
	}
	\centering
	\vspace{-1pt}
	\captionsetup[subfloat]{position=top,captionskip=-30pt}
	\subfloat[][]{\label{fig_ageing_RADP13_P1000B}
		\centering
		\includegraphics[trim={-18ex 10ex 0 10ex},scale=0.265,clip]{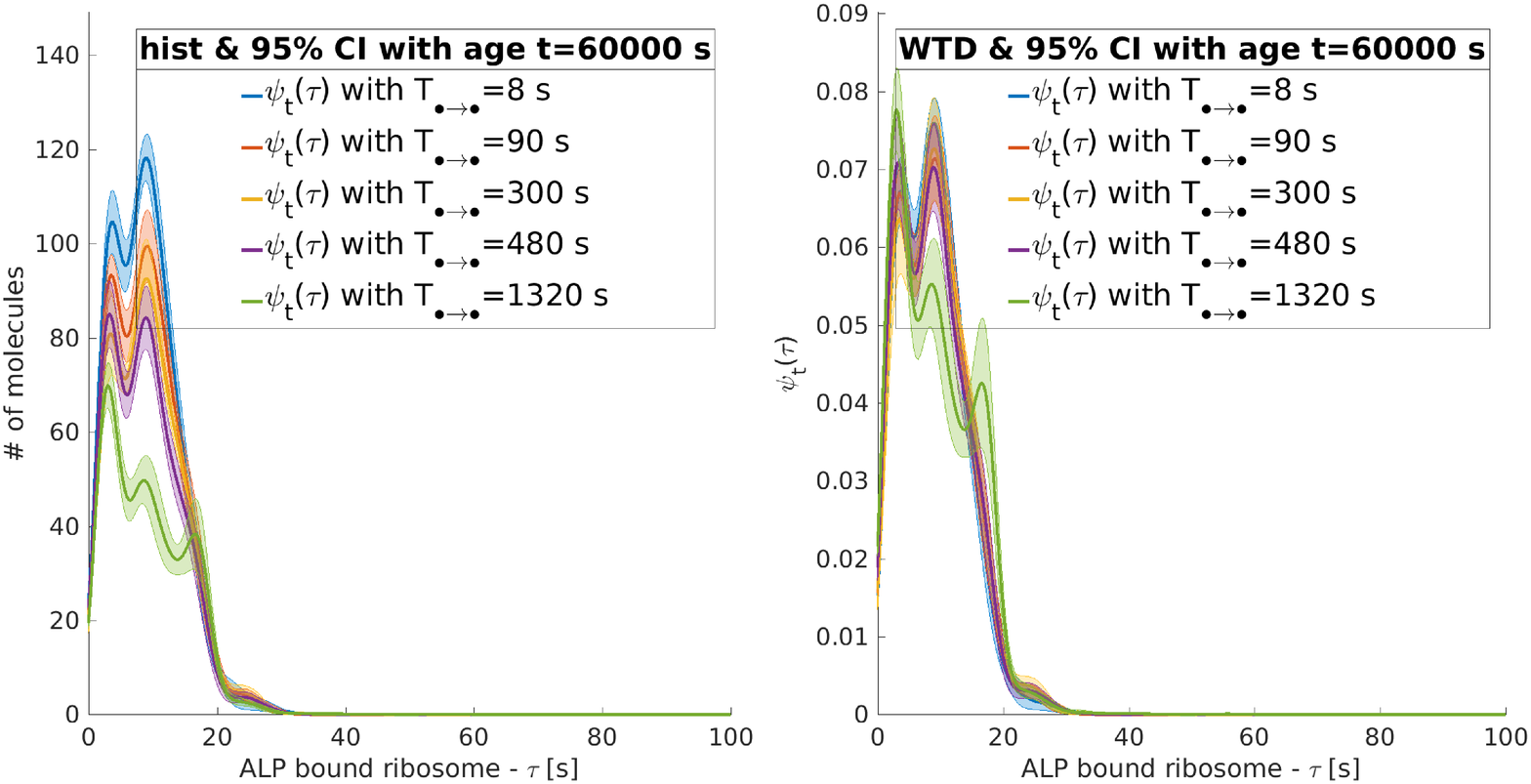}
	}
	\caption{\csentence{Ageing of ALP mRNA.} 
		For $\textrm{\RADP13}=1320\,s$, the WTDs of the ALP mRNA have trimodal distributions with peaks at $4\,s$, $10\,s$ and $18\,s$. The three modes are persistent  at any age of the system.  For $\textrm{\RADP13}\leq 480\,s$, the WTDs' slope at  $\tau>10\,s$ remains negative and the shape of the WTDs is bimodal. 
		\textbf{\protect\subref{fig_ageing_RADP13_P1000A}} and \textbf{\protect\subref{fig_ageing_RADP13_P1000B}} $M=10000\,\mu Pa$ and $P=1000\,s$,
		while \RADP14 changes; \textbf{\protect\subref{fig_ageing_RADP13_P1000B}} aged condition of the case on the above line.
	}
	\label{fig_ageing_RADP13_P1000}		
\end{figure}

\begin{figure}[h!]
	\centering
	\captionsetup[subfloat]{position=top,captionskip=-18pt}
	\subfloat[][]{\label{fig_ageing_RADP1113_P200000C}
		\centering
		\includegraphics[trim={-18ex 10ex 0 27ex},scale=0.265,clip]{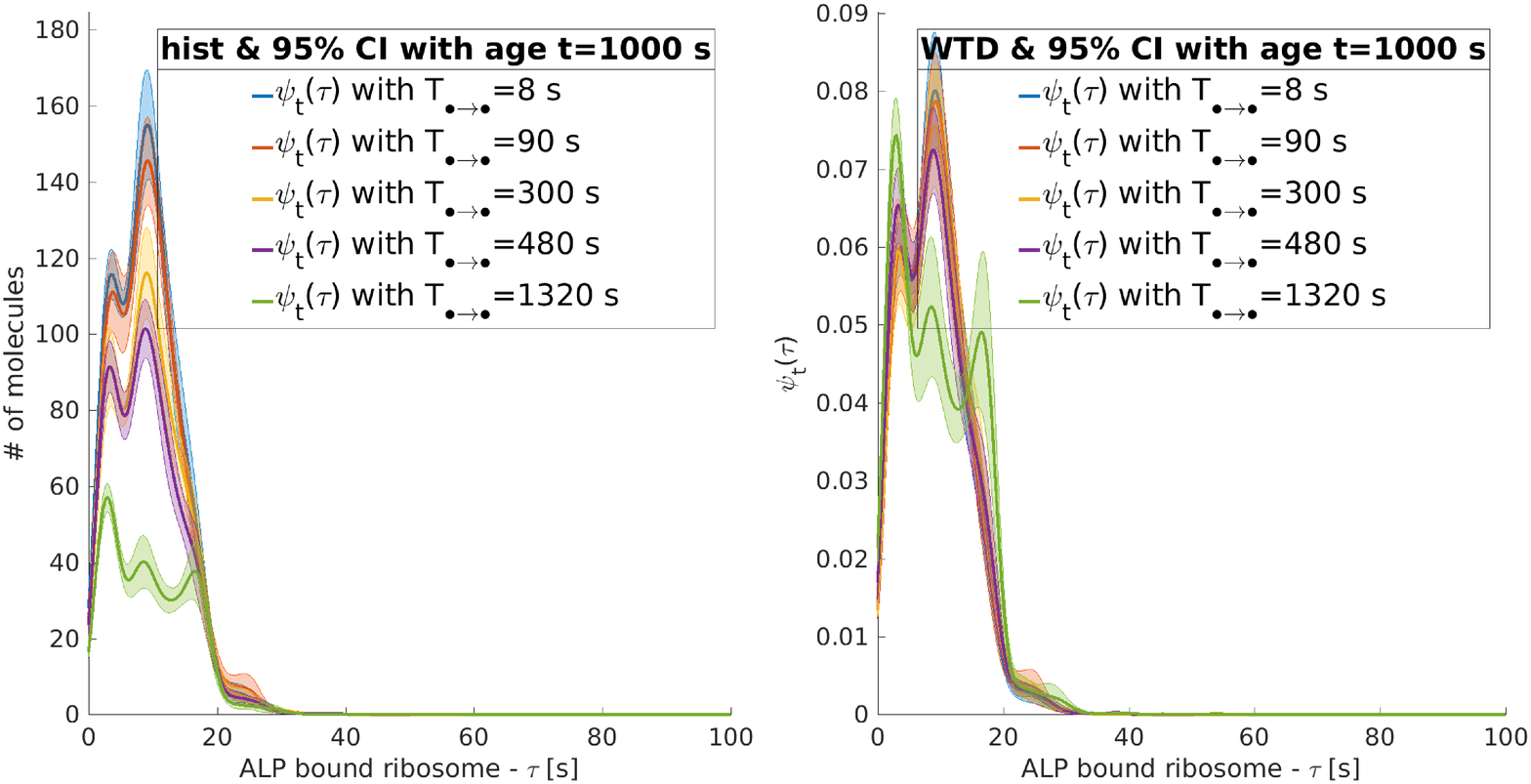}
	}
	\vspace{-1pt}
	\captionsetup[subfloat]{position=top,captionskip=-18pt}
	\subfloat[][]{\label{fig_ageing_RADP1113_P200000D}
		\centering
		\includegraphics[trim={-18ex 10ex 0 27ex},scale=0.265,clip]{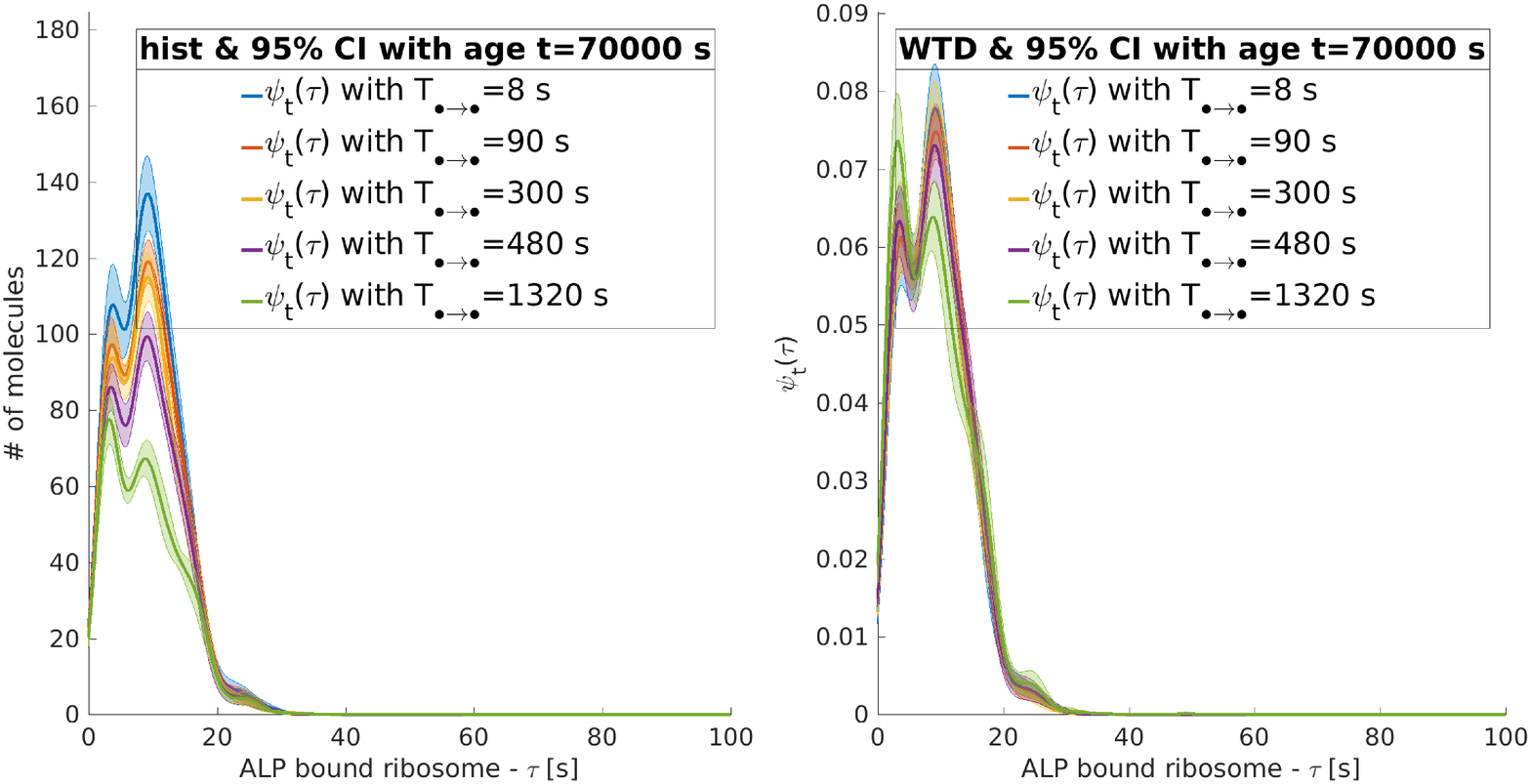}
	}
	\vspace{-1pt}
	\captionsetup[subfloat]{position=top,captionskip=-18pt}
	\subfloat[][]{\label{fig_ageing_RADP1113_P200000A}
		\centering
		\includegraphics[trim={-18ex 10ex 0 27ex},scale=0.265,clip]{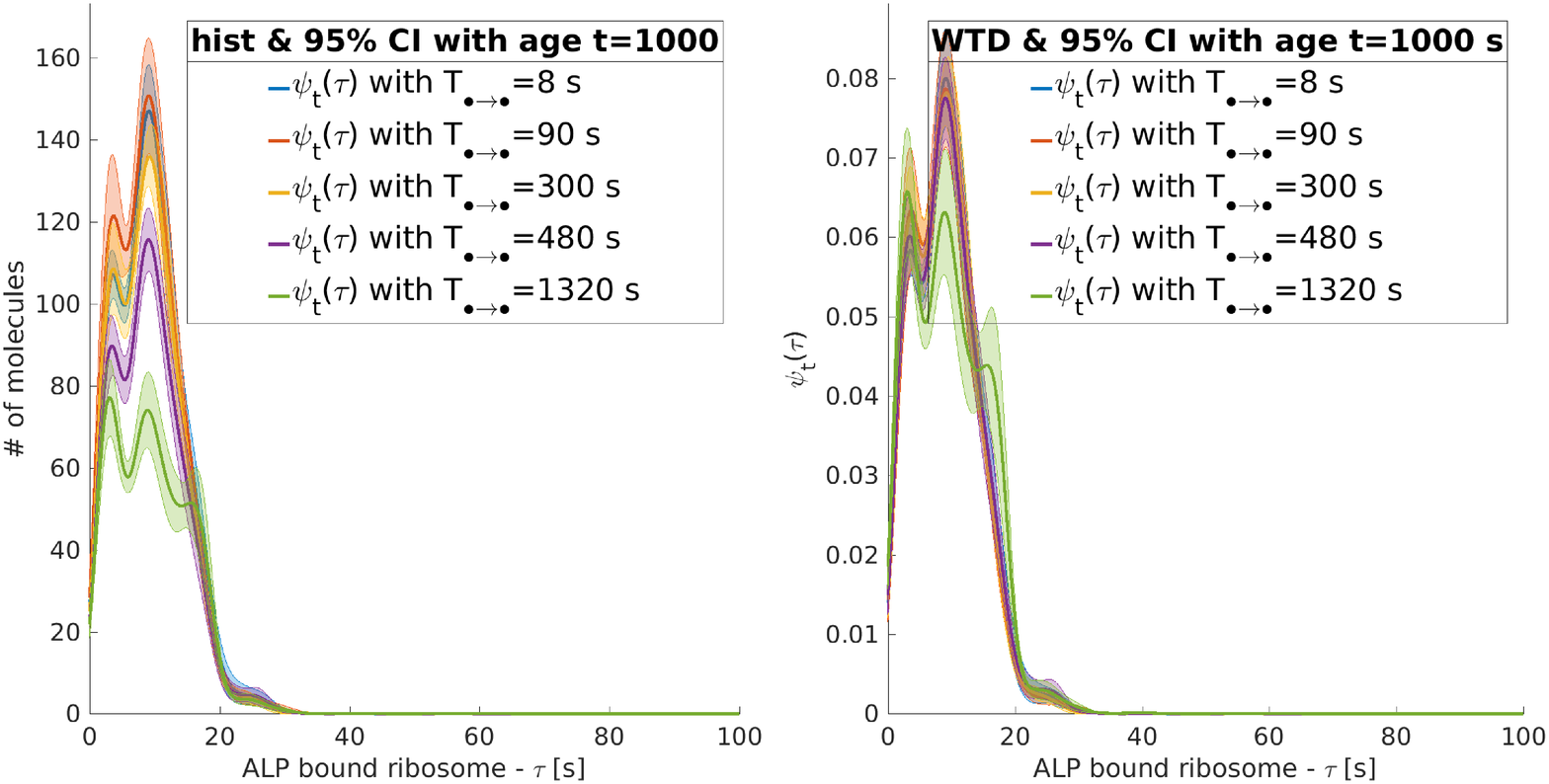}
	}
	\vspace{-1pt}
	\captionsetup[subfloat]{position=top,captionskip=-18pt}
	\subfloat[][]{\label{fig_ageing_RADP1113_P200000B}
		\centering
		\includegraphics[trim={-18ex 10ex 0 27ex},scale=0.265,clip]{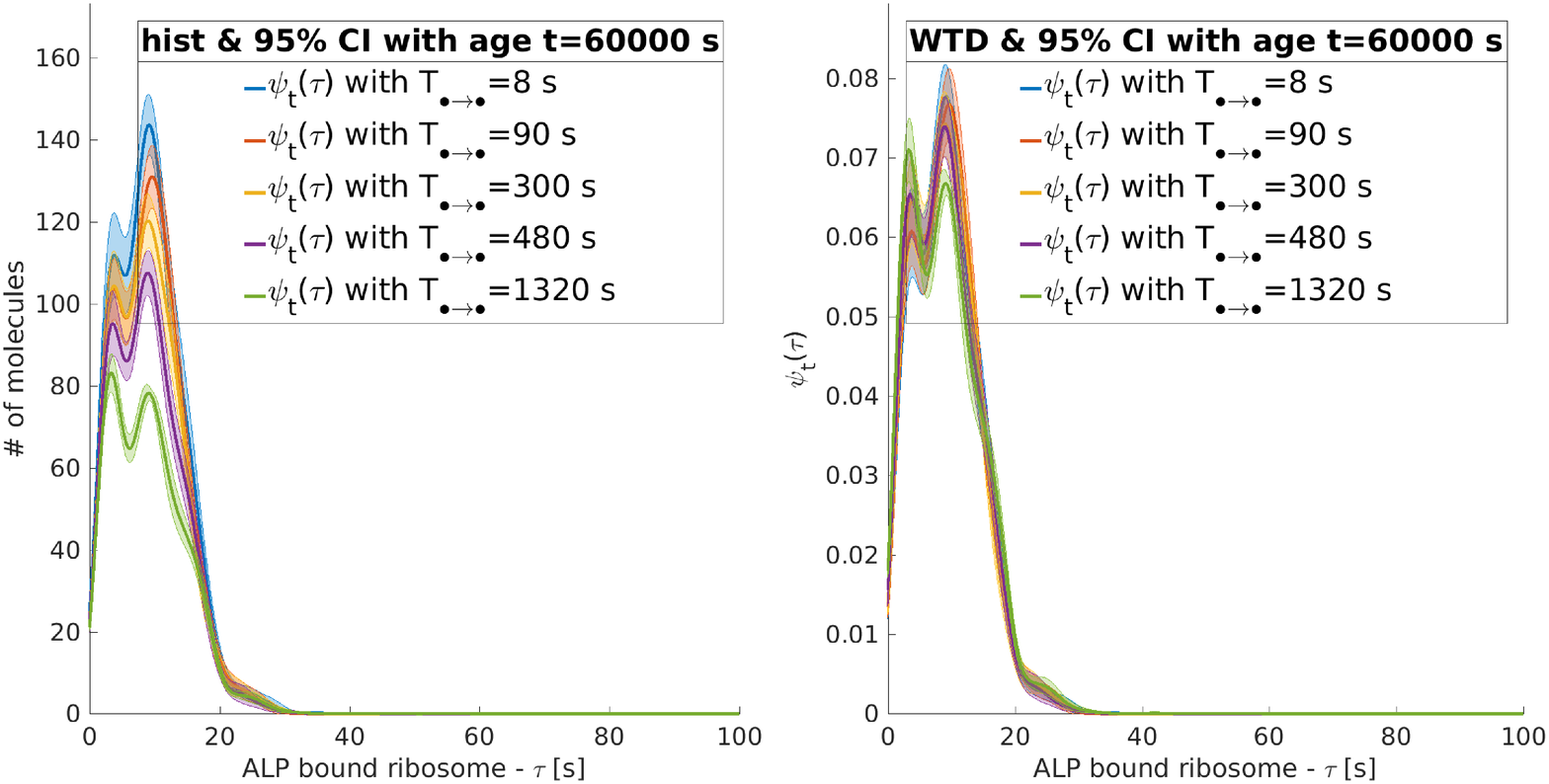}
	}
	\caption{\csentence{Ageing of ALP mRNA.} 
		For $\textrm{\RADP11}=$ $\textrm{\RADP13}$$=1320\,s$, the WTDs of the ALP mRNA at $t=0\,s$ have trimodal distributions with peaks at $4\,s$, $10\,s$ and $18\,s$. After the system reaches age $t=5000\,s$, the WTDs' slope at  $\tau>10\,s$ remains negative. In the other cases, the peak at $\tau=18\,s$ is not present. 
		\textbf{\protect\subref{fig_ageing_RADP1113_P200000C}}-\textbf{\protect\subref{fig_ageing_RADP1113_P200000B}} $M=10000\,\mu Pa$ and $P=200000$;
		\textbf{\protect\subref{fig_ageing_RADP1113_P200000C}}	\RADP13 changes; \textbf{\protect\subref{fig_ageing_RADP1113_P200000D}} aged condition of the case on the above line;
		\textbf{\protect\subref{fig_ageing_RADP1113_P200000A}}	\RADP11 changes; \textbf{\protect\subref{fig_ageing_RADP1113_P200000B}} aged condition of the case on the above line. 
	}
	\label{fig_ageing_RADP1113_P200000}	
\end{figure}

\begin{figure}[h!]
	\centering
	\captionsetup[subfloat]{position=top,captionskip=-30pt}
	\subfloat[][]{\label{fig_free_mRNAA}
		\centering
		\includegraphics[trim={-18ex 10ex 0 10ex},scale=0.265,clip]{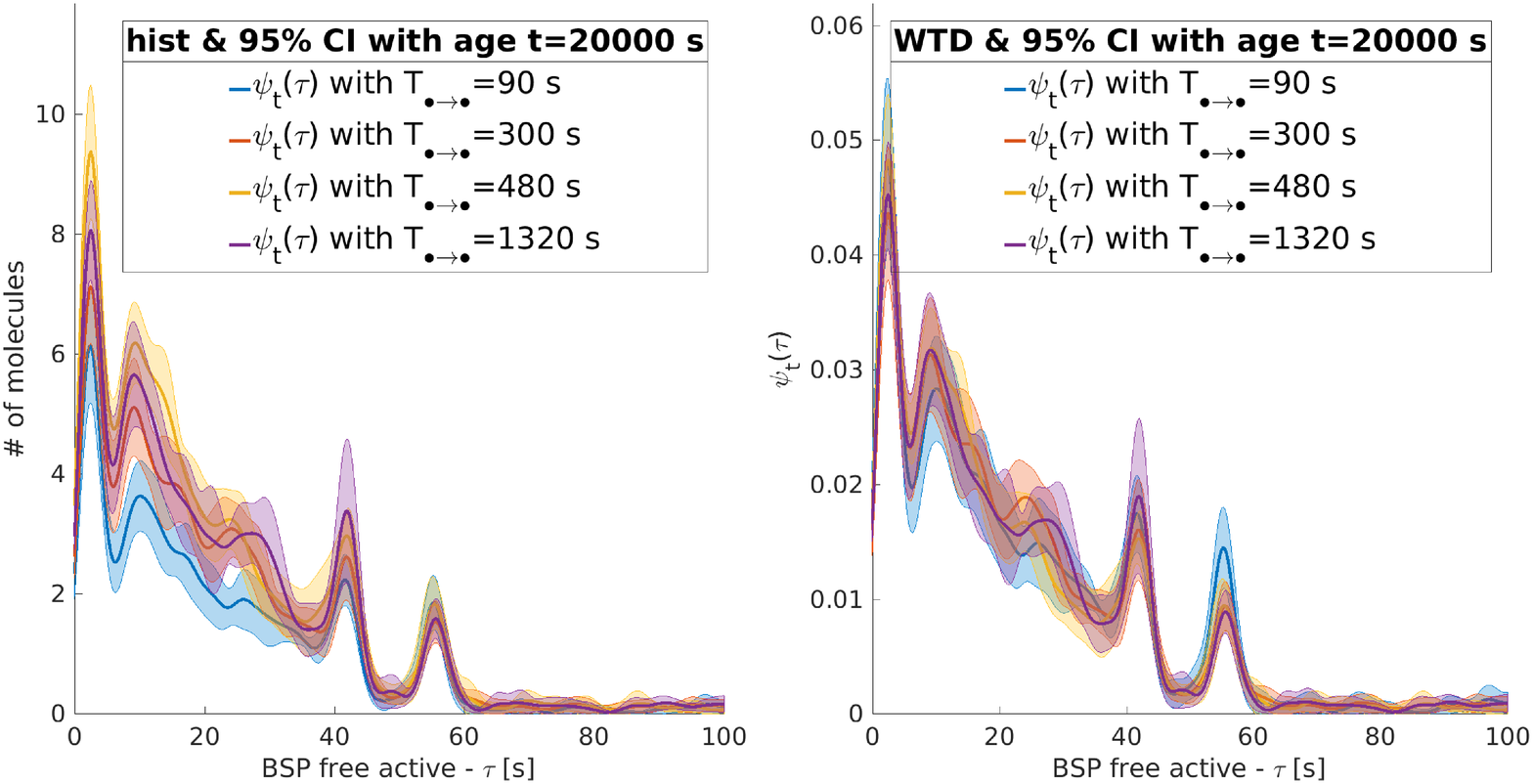}
	}
	\centering
	\vspace{-1pt}
	\captionsetup[subfloat]{position=top,captionskip=-30pt}
	\subfloat[][]{\label{fig_free_mRNAB}
		\centering
		\includegraphics[trim={-18ex 10ex 0 10ex},scale=0.265,clip]{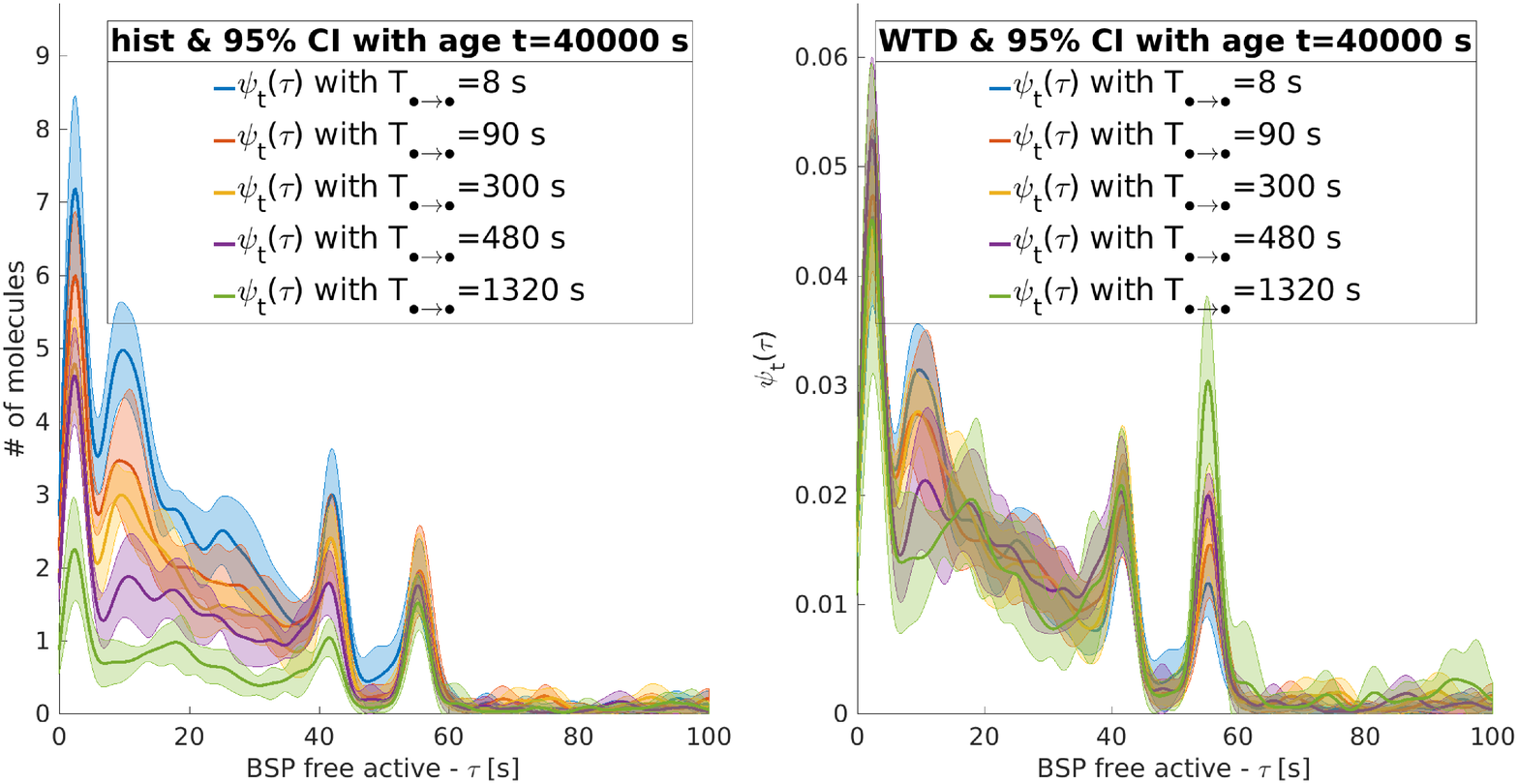}
	}
	\caption{\csentence{BSP mRNA.} 
		The BSP mRNA WTDs have 4 modes. The shape of the distributions are similar to all the unbounded mRNA in the cytoplasm. The modes are at $\tau$ equal to $5\,s$, $10\,s$, $41\,s$ and $55\,s$. In the region between the modes at $10\,s$ and $41\,s$, the distributions are noisy and their average slope depend on $T_{\bullet\ \rightarrow\ \bullet}$. \textbf{\protect\subref{fig_free_mRNAA}}  $P=1000\,s$, $M=10000\,\mu Pa$ and \RADP12 change; \textbf{\protect\subref{fig_free_mRNAB}} $P=200000\,s$, $M=10000\,\mu Pa$ and \RADP13 change.
	}
	\label{fig_free_mRNA}
\end{figure}

\begin{figure}[h!]
	\captionsetup[subfloat]{position=top,captionskip=-18pt}
	\subfloat[][]{\label{fig_RADP14t1000editedA}
		\centering
		\includegraphics[trim={-18ex 10ex 100ex 0},scale=0.265,clip]{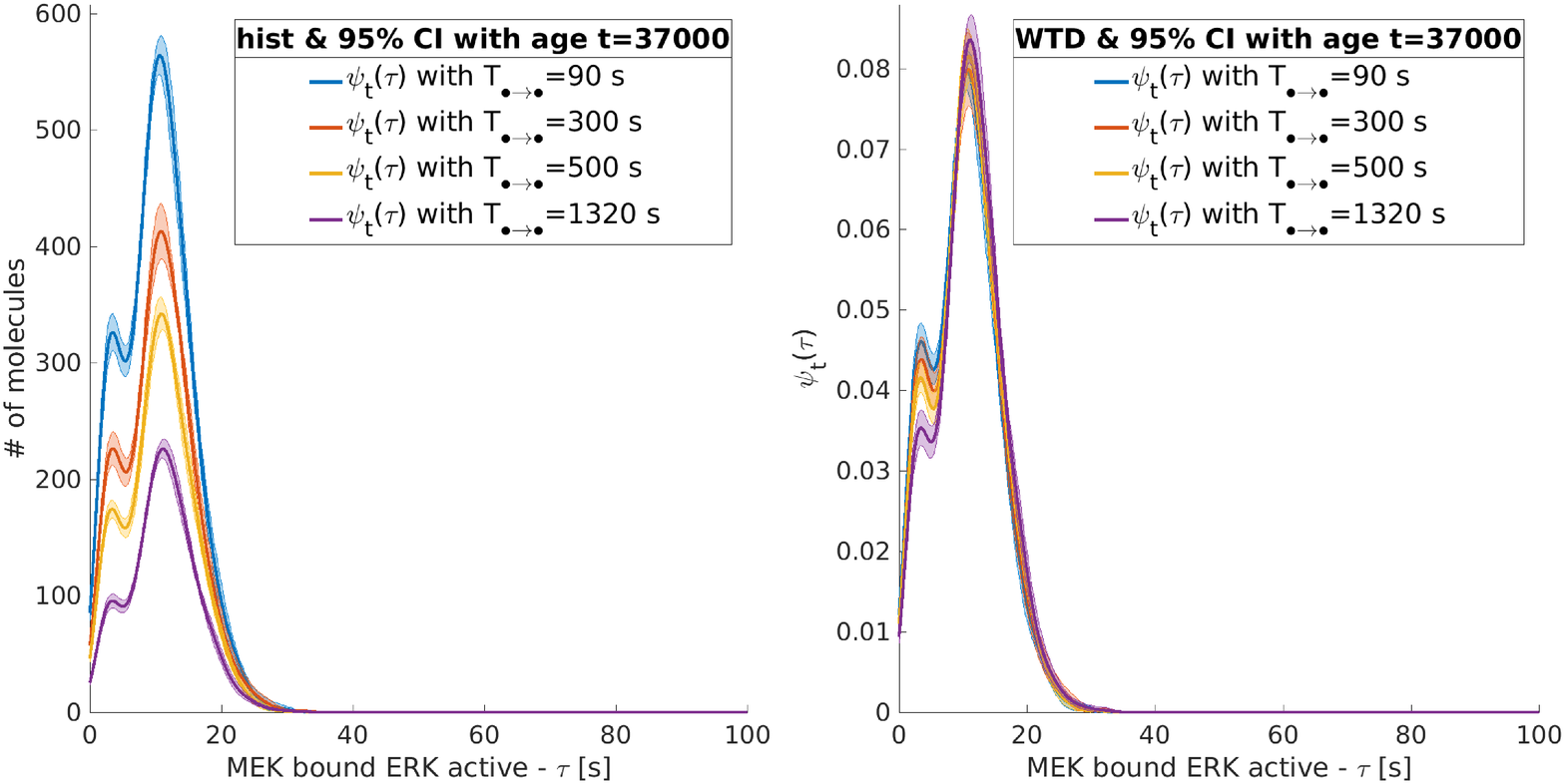}
	}
	\vspace{-1pt}
	\captionsetup[subfloat]{position=top,captionskip=-18pt}
	\subfloat[][]{\label{fig_RADP14t1000editedB}
		\centering
		\includegraphics[trim={-18ex 10ex 100ex 0},scale=0.265,clip]{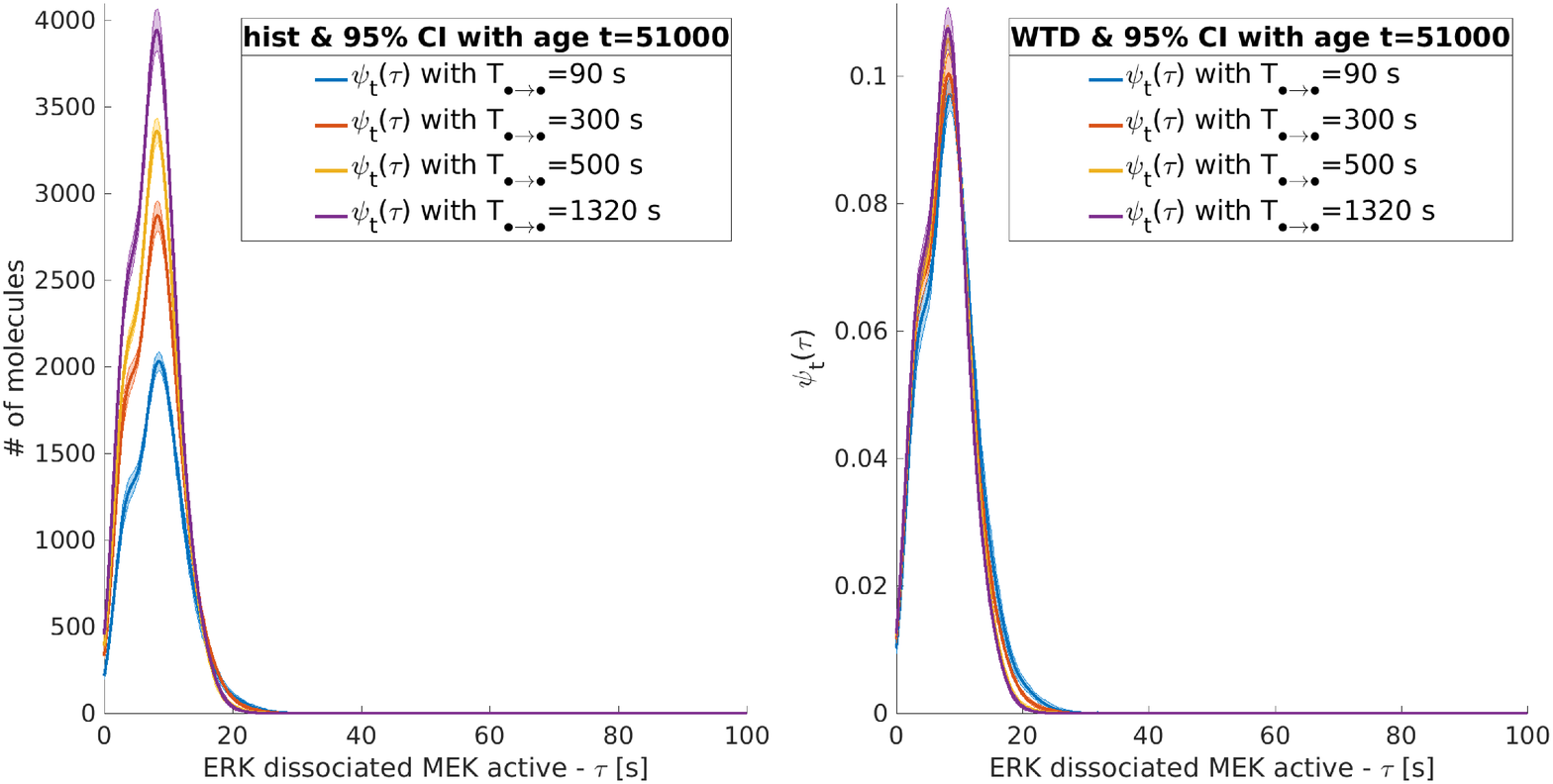}
	}
	\vspace{-1pt}
	\captionsetup[subfloat]{position=top,captionskip=-18pt}
	\subfloat[][]{\label{fig_RADP14t1000editedC}
		\centering
		\includegraphics[trim={-18ex 10ex 100ex 0},scale=0.265,clip]{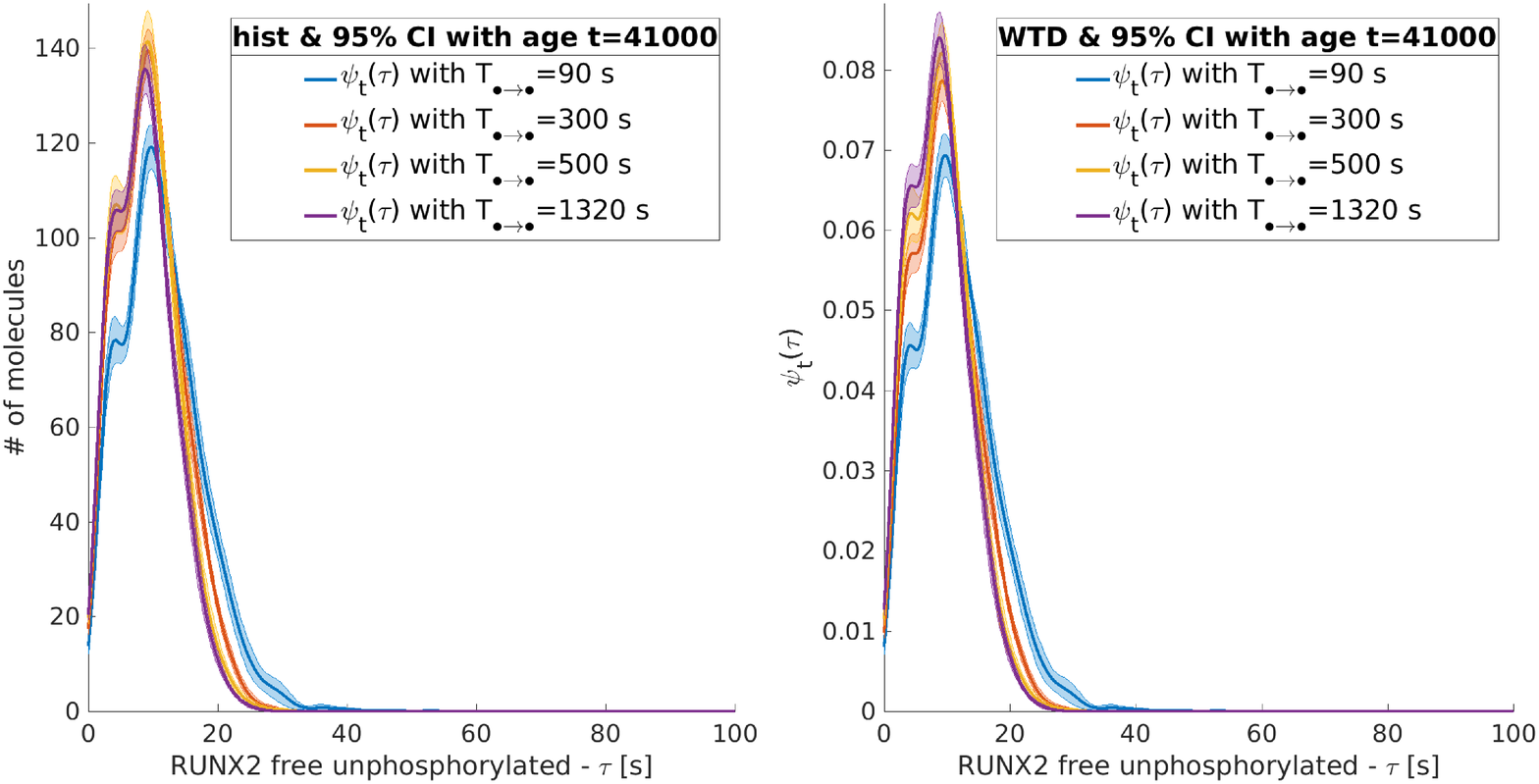}
	}
	\caption{\csentence{MEK+ERK, ERK and RUNX2.} 
		Distributions for different values of \RADP14 under a periodic perturbation with $P=1000\,s$ and $\M=10000\,\mu Pa$. \RADP14 produces larger effects on the next nearest neighbour of the network like unphosphorylated RUNX2 and unphosphorylated ERK. Variations in the distributions and in the non-normalized histogram of MEK bounded to ERK. No effects are visible on the distribution of active ERK dissociated from MEK. Amplitudes or number of fluctuations of ERK dissociated MEK increase with the increase of \RADP14, while they show an inverse relation for MEK bound to ERK decrease.  \textbf{\protect\subref{fig_RADP14t1000editedA}} $\MEK$ bound to $\ERK$;
		\textbf{\protect\subref{fig_RADP14t1000editedB}} active $\ERK$ dissociated from $\MEK$; \textbf{\protect\subref{fig_RADP14t1000editedC}} unphosphorylated $\RUNX$.
	}
	\label{fig_RADP14t1000edited}
	
\end{figure}

\begin{figure}[h!]
	\captionsetup[subfloat]{position=top,captionskip=-18pt}
	\subfloat[][]{\label{fig_RADP14t200000editedA}
		\centering
		\includegraphics[trim={-18ex 10ex 100ex 0},scale=0.265,clip]{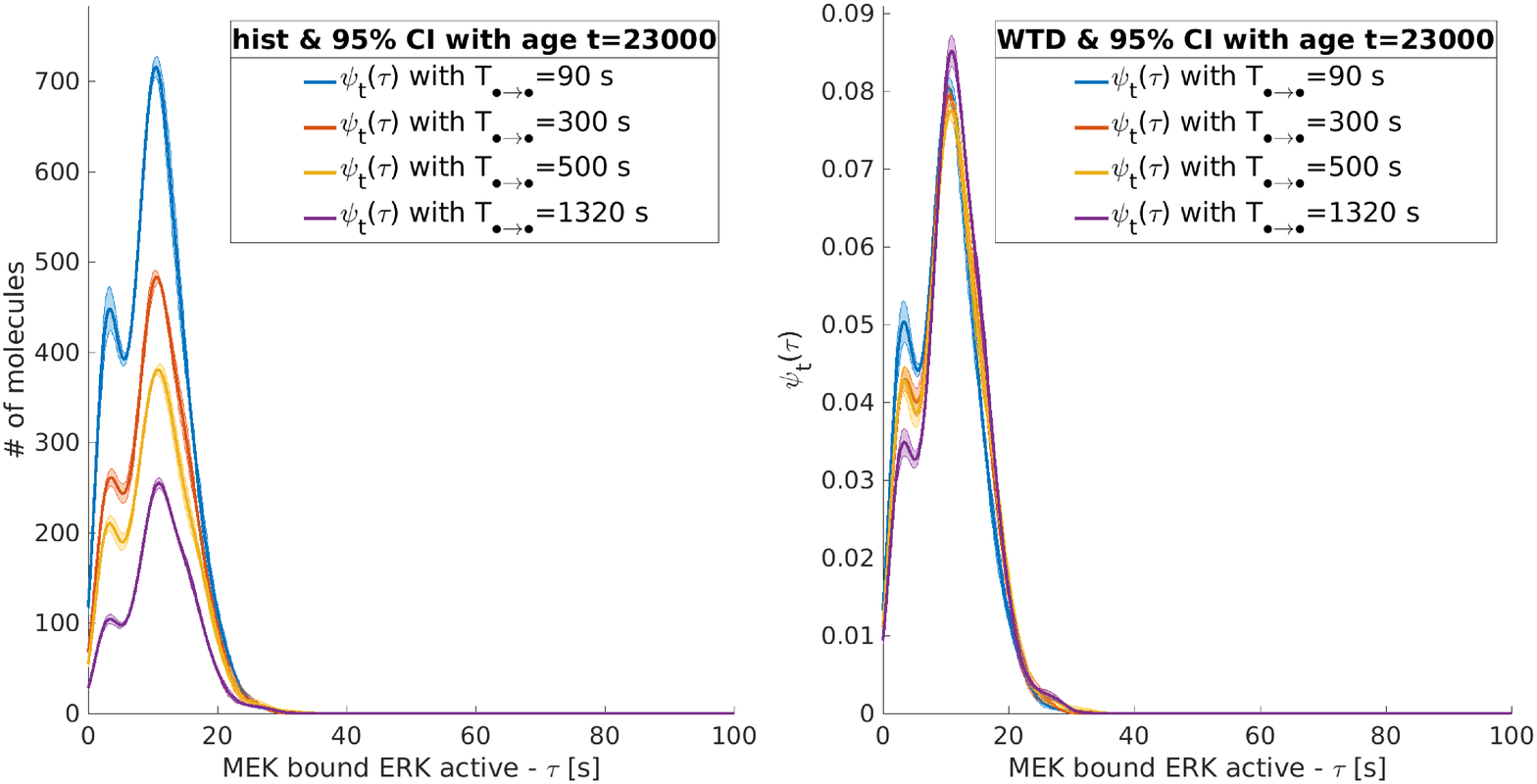}
	}
	\vspace{-1pt}
	\captionsetup[subfloat]{position=top,captionskip=-18pt}
	\subfloat[][]{\label{fig_RADP14t200000editedB}
		\centering
		\includegraphics[trim={-18ex 10ex 100ex 0},scale=0.265,clip]{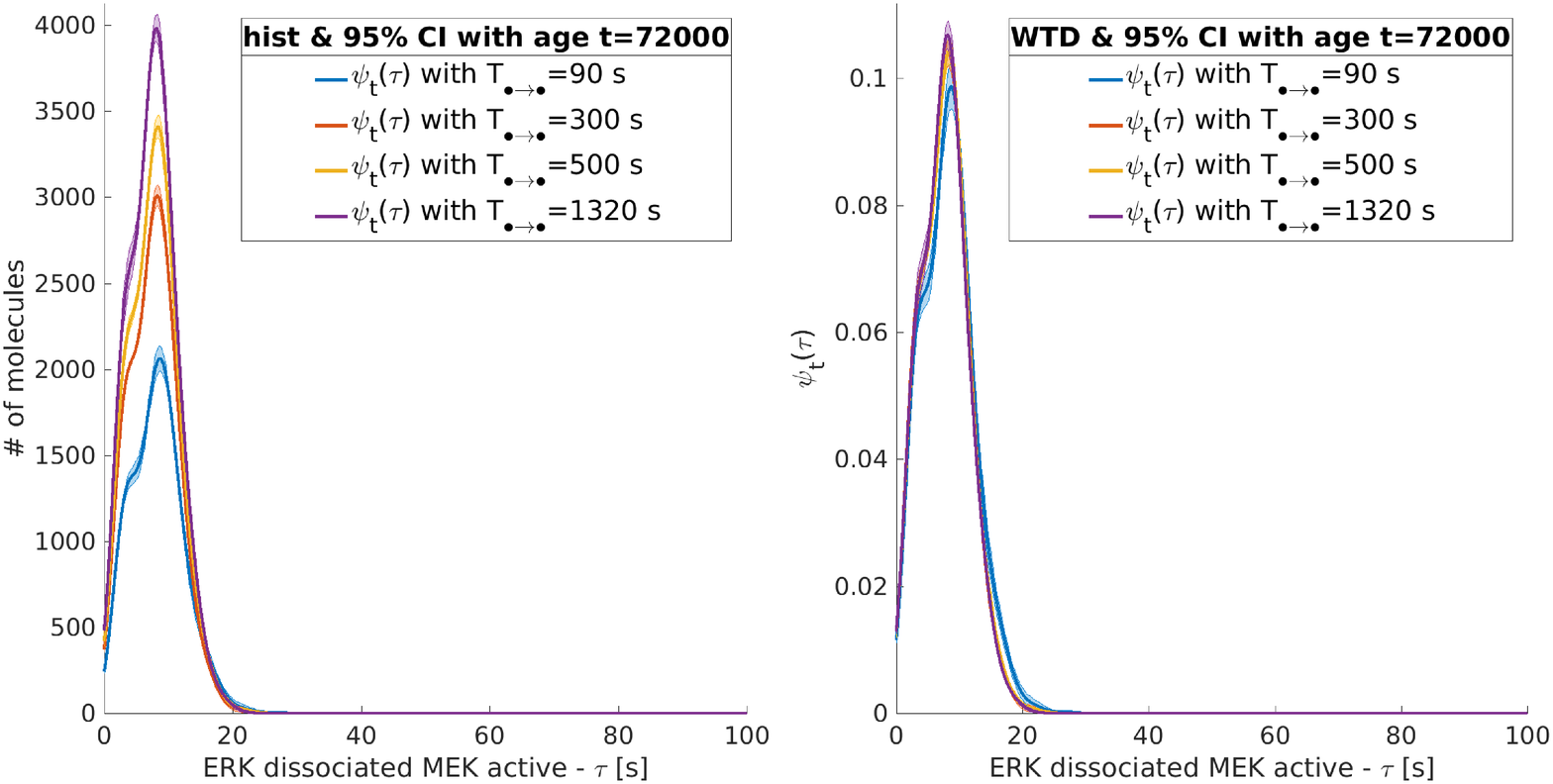}
	}
	\vspace{-1pt}
	\captionsetup[subfloat]{position=top,captionskip=-18pt}
	\subfloat[][]{\label{fig_RADP14t200000editedC}
		\centering
		\includegraphics[trim={-18ex 10ex 100ex 0},scale=0.265,clip]{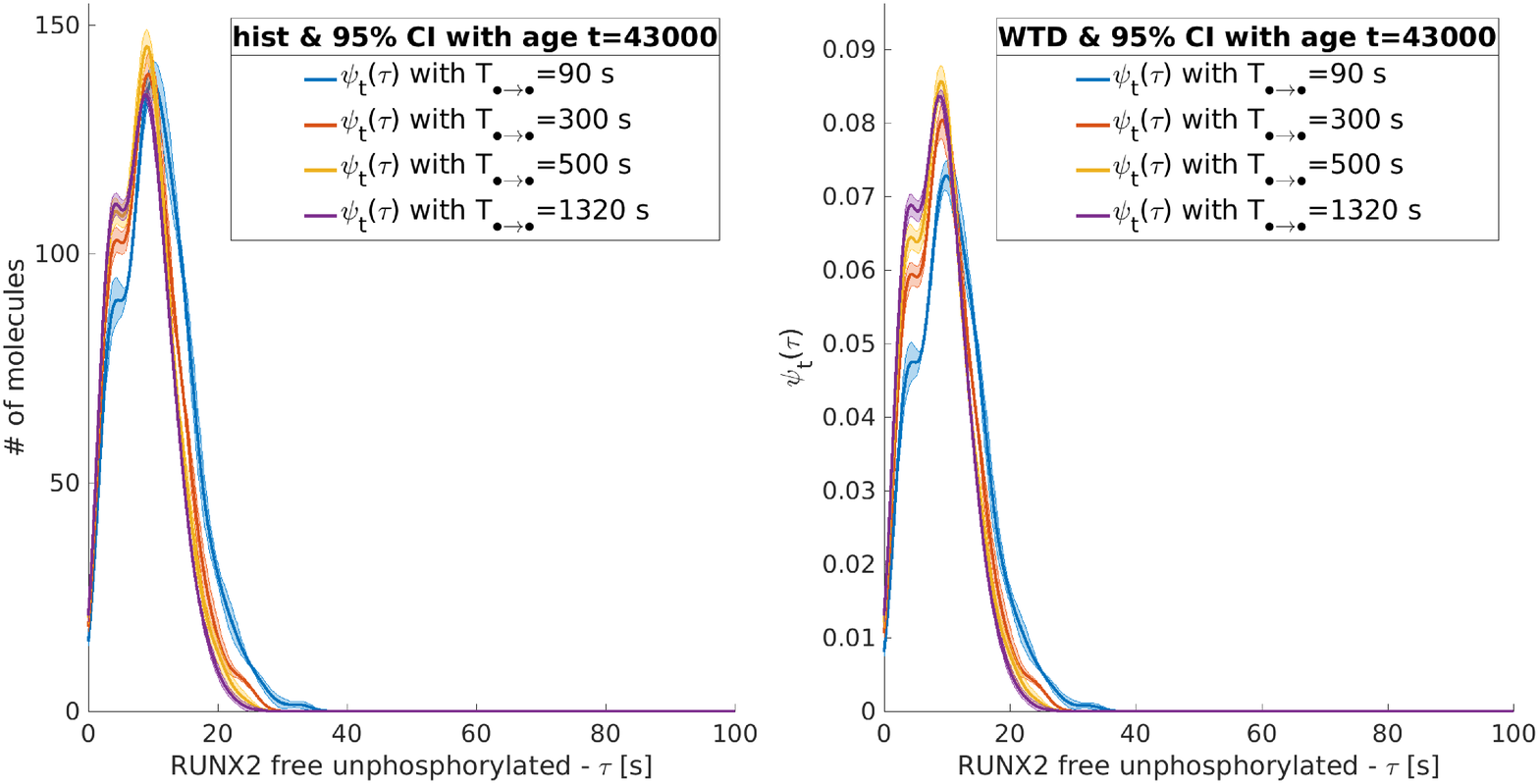}
	}
	\caption{\csentence{MEK+ERK, ERK and RUNX2.} 
		Distributions for different values of \RADP14 under a periodic perturbation with $P=200000\,s$ and $\M=10000\,\mu Pa$. The distributions are similar to those obtained with higher frequency  stimulation (see \fref{fig_RADP14t1000edited}). The histograms show larger or more frequent fluctuations of unphosphorylated  RUNX2 and MEK bound ERK.
\textbf{\protect\subref{fig_RADP14t200000editedA}} $\MEK$ bound to $\ERK$;
\textbf{\protect\subref{fig_RADP14t200000editedB}} active $\ERK$ dissociated from $\MEK$; \textbf{\protect\subref{fig_RADP14t200000editedC}} unphosphorylated $\RUNX$.	
}
	\label{fig_RADP14t200000edited}
\end{figure}

\begin{figure}[h!]
	\centering
	\captionsetup[subfloat]{position=top,captionskip=-30pt}
	\subfloat[][]{\label{fig_saturationA}
		\centering
		\includegraphics[trim={-18ex 10ex 0 10ex},scale=0.265,clip]{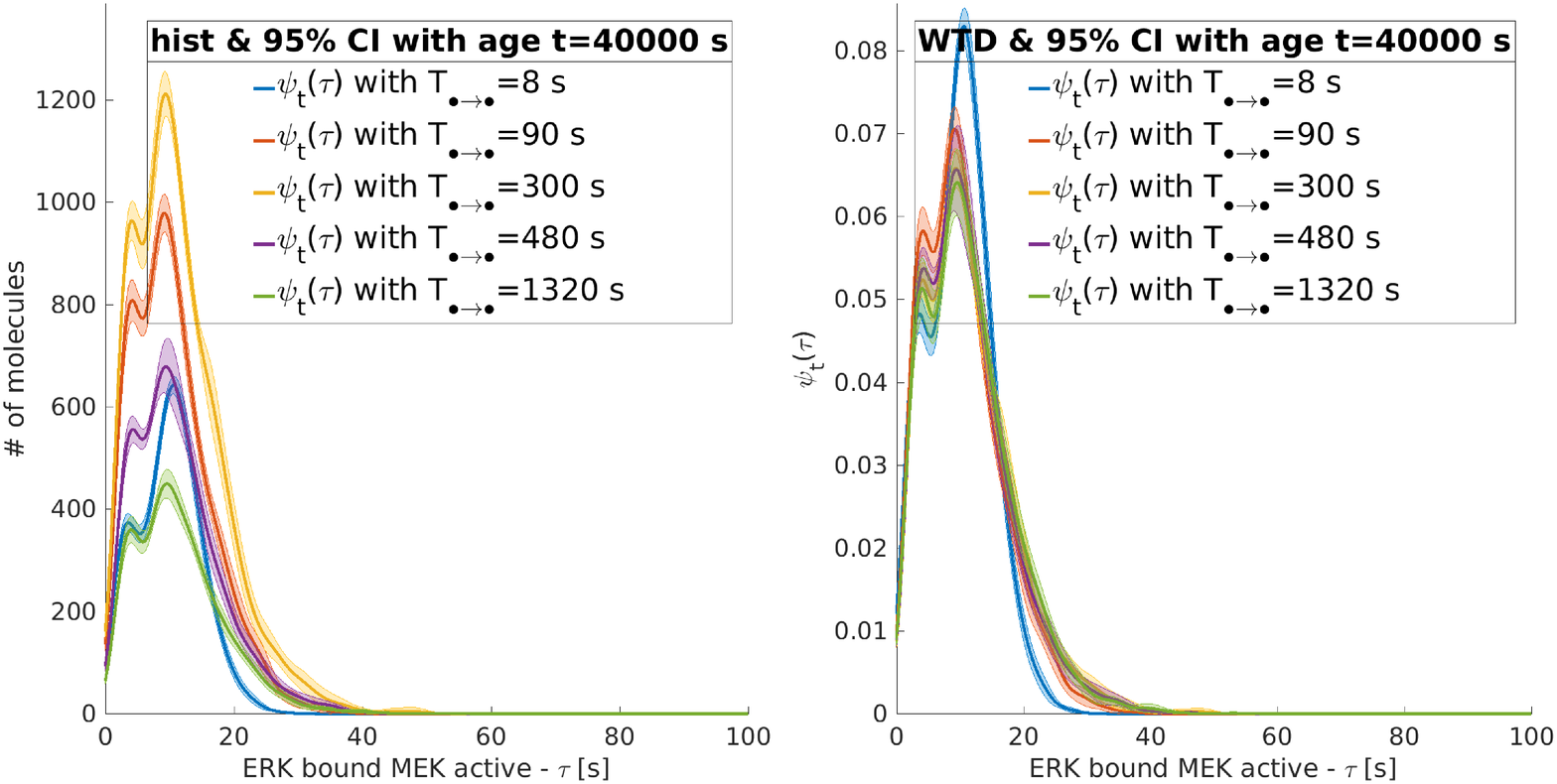}
	}
	\vspace{-1pt}
	\captionsetup[subfloat]{position=top,captionskip=-30pt}
	\subfloat[][]{\label{fig_saturationB}
		\centering
		\includegraphics[trim={-18ex 10ex 0 10ex},scale=0.265,clip]{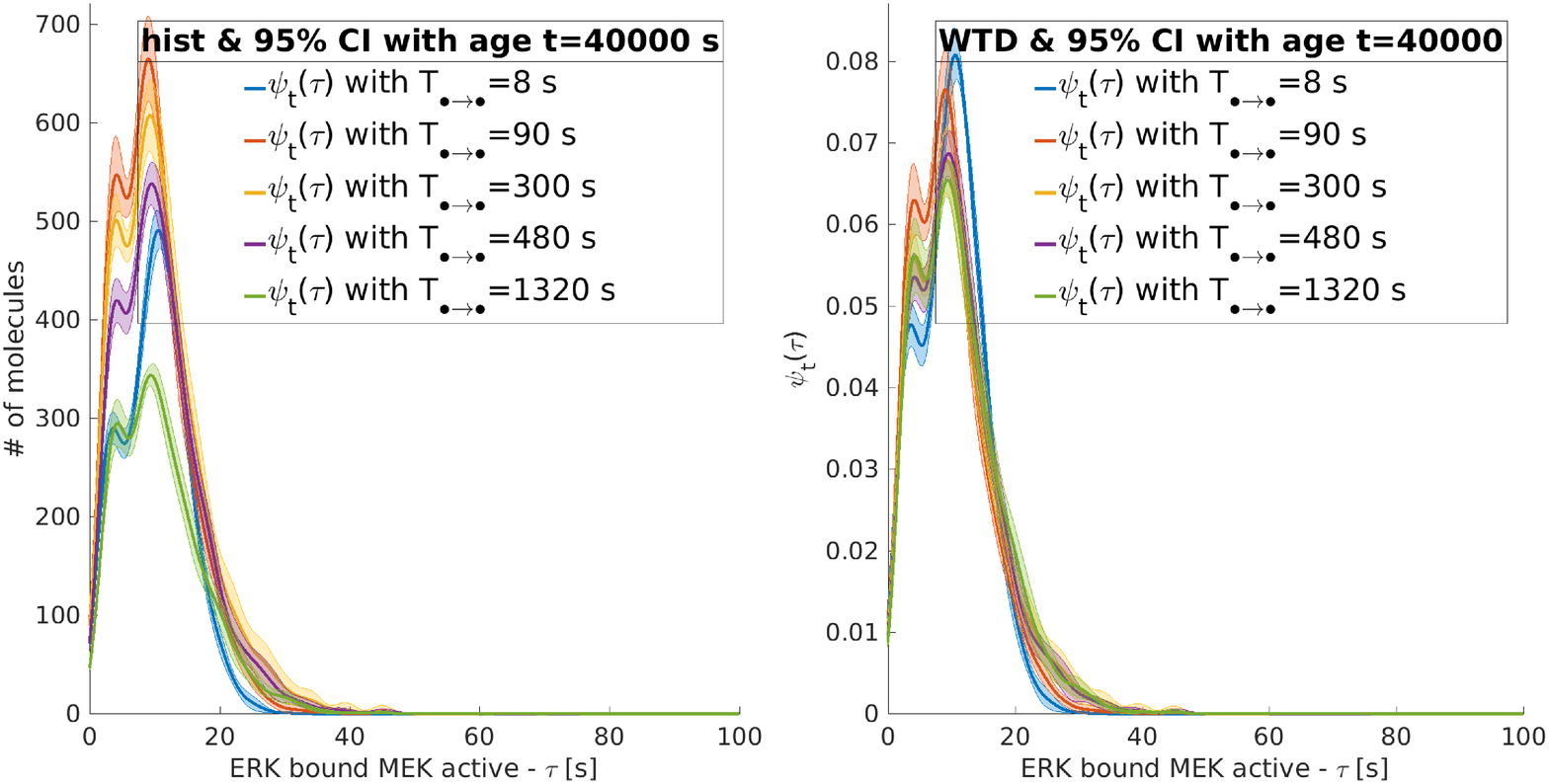}
	}
	\vspace{-1pt}
	\captionsetup[subfloat]{position=top,captionskip=-30pt}
	\subfloat[][]{\label{fig_saturationC}
		\centering
		\includegraphics[trim={-18ex 10ex 0 10ex},scale=0.265,clip]{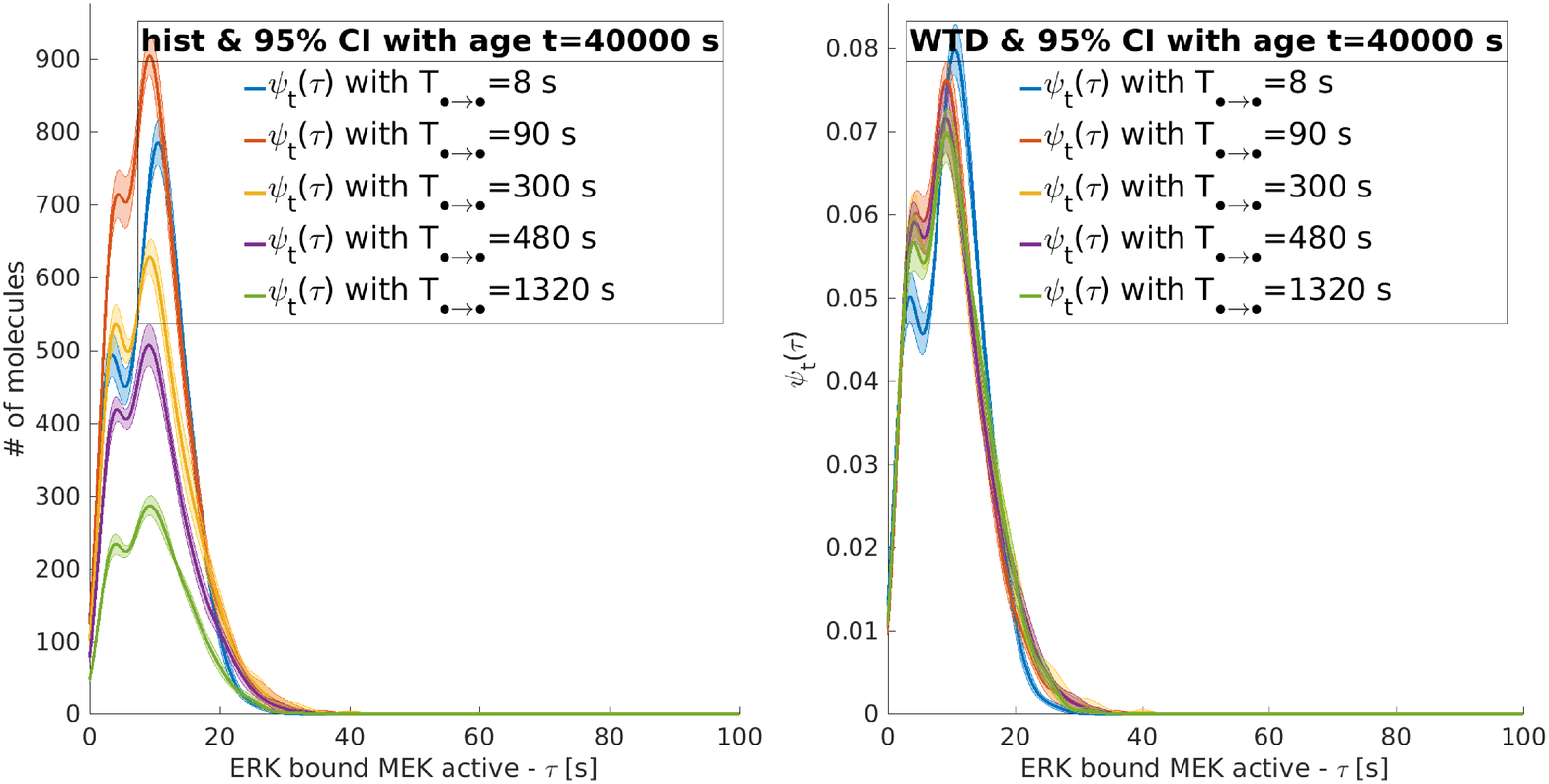}
	}
	\caption{\csentence{ERK+MEK complex.}
		The distributions of intertimes for the ERK+MEK complex WTDs are bimodal. The modes are at $\tau$ equal to $4\,s$ and $15\,s$. When $\textrm{\RADP13}=8\,s$, the mode at $\tau=15\,s$ shifts to $17\,s$. The non normalized histograms on the left column show that the total number of events is at the minimum when \RADP13 has the maximum value simulated equal to $1320s\,$. The number of critical events increases  when \RADP13 decreases and reaches a critical value after which the dissociation time is directly proportional to the number of events.  \textbf{\protect\subref{fig_saturationA}}-\textbf{\protect\subref{fig_saturationC}} $M=10000\,\mu Pa$; \textbf{\protect\subref{fig_saturationA}} $P=1000\,s$; \textbf{\protect\subref{fig_saturationB}} $P=5000\,s$;
		\textbf{\protect\subref{fig_saturationC}} $P=200000\,s$.
	}
	\label{fig_saturation}
\end{figure}

\begin{figure}[h!]
	\centering
	\captionsetup[subfloat]{position=top,captionskip=-28pt}
	\subfloat[][]{\label{fig_noiseA}
		\centering
		\includegraphics[trim={-18ex 10ex 0 10ex},scale=0.265,clip]{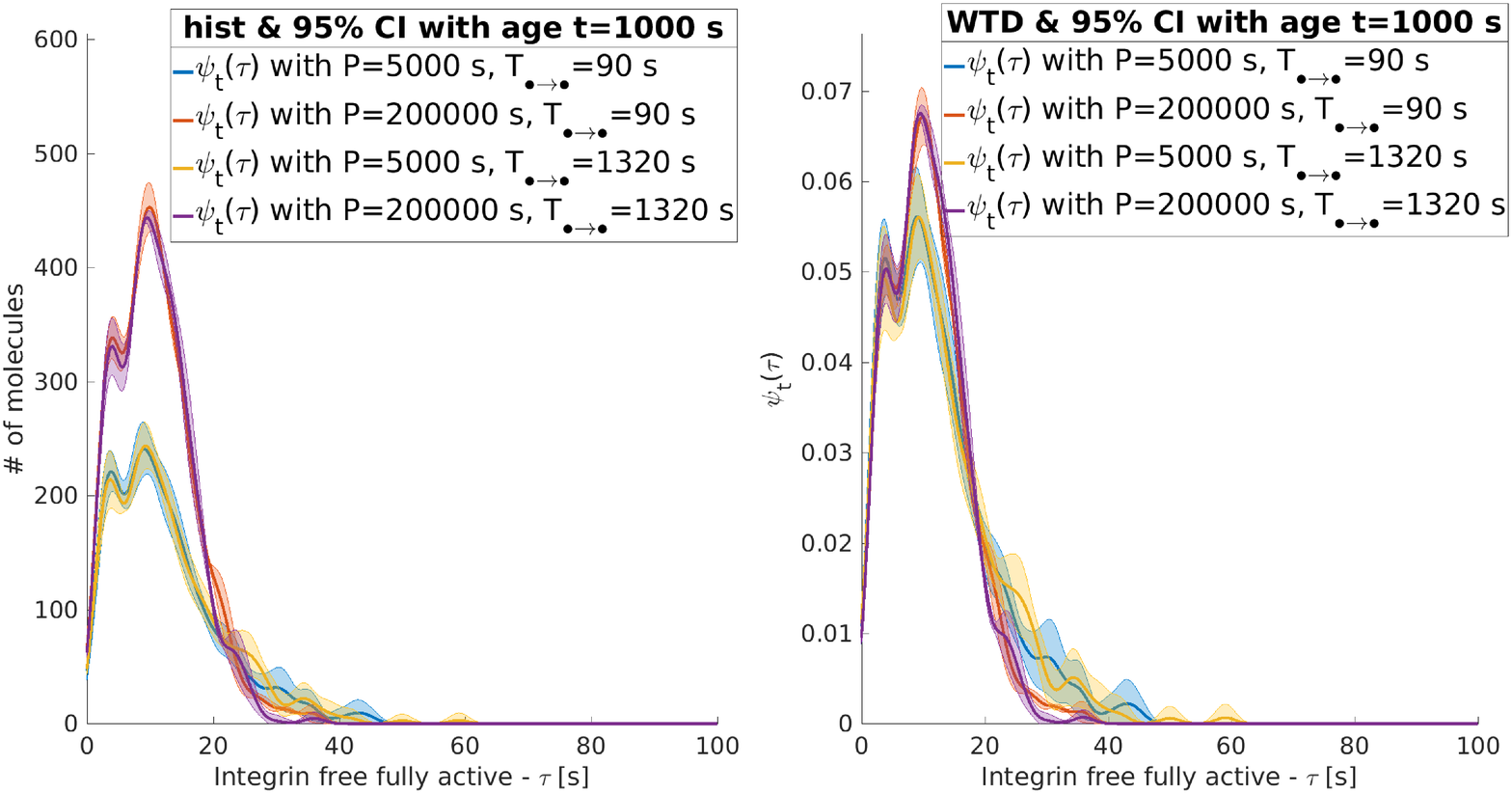}
	}
	\vspace{-4pt}
	\captionsetup[subfloat]{position=top,captionskip=-28pt}
	\subfloat[][]{\label{fig_noiseB}
		\centering
		\includegraphics[trim={-18ex 10ex 0 10ex},scale=0.265,clip]{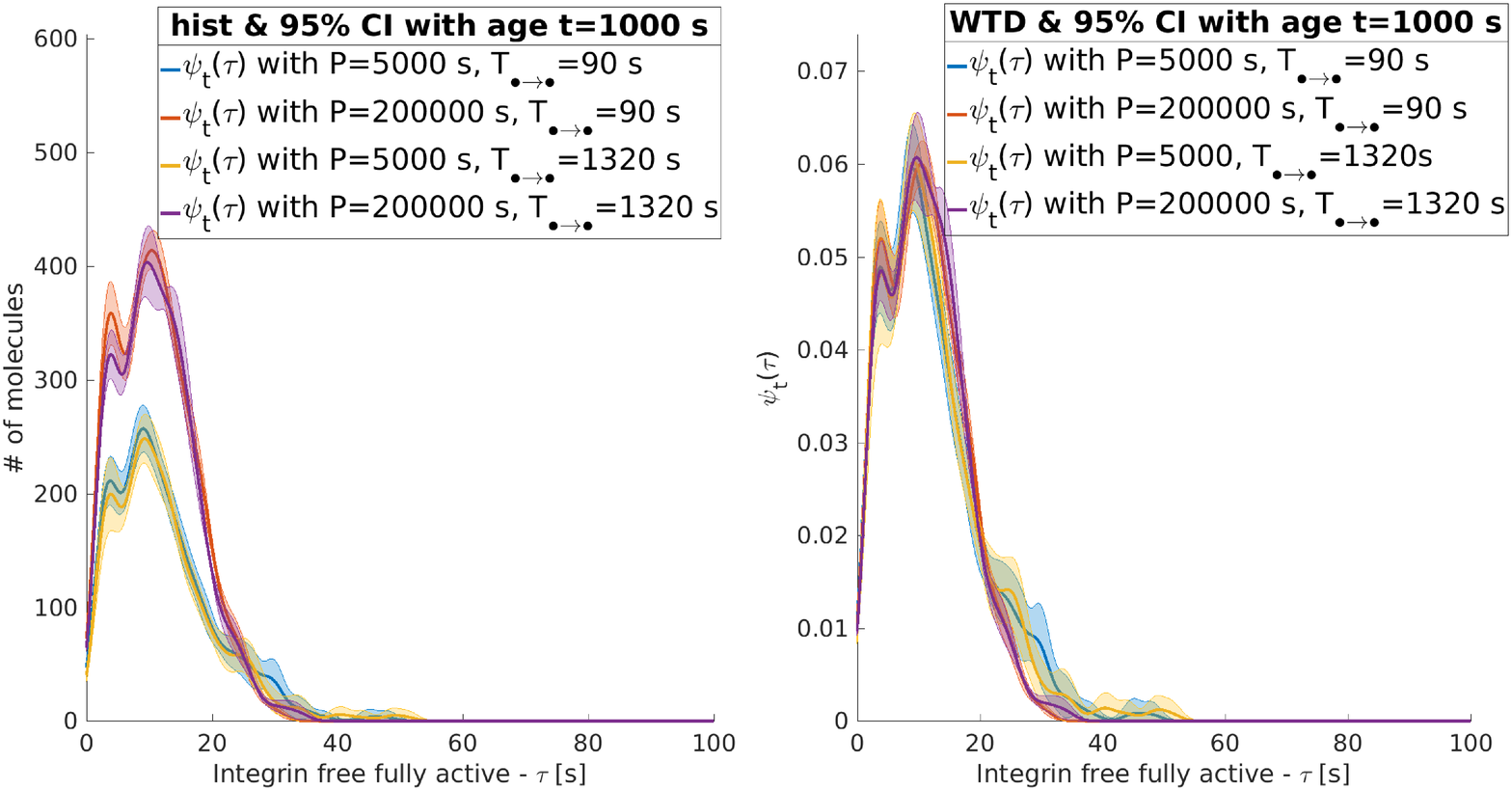}
	}
	\vspace{-4pt}
	\captionsetup[subfloat]{position=top,captionskip=-28pt}
	\subfloat[][]{\label{fig_noiseC}
		\centering
		\includegraphics[trim={-18ex 10ex 0ex 10ex},scale=0.265,clip]{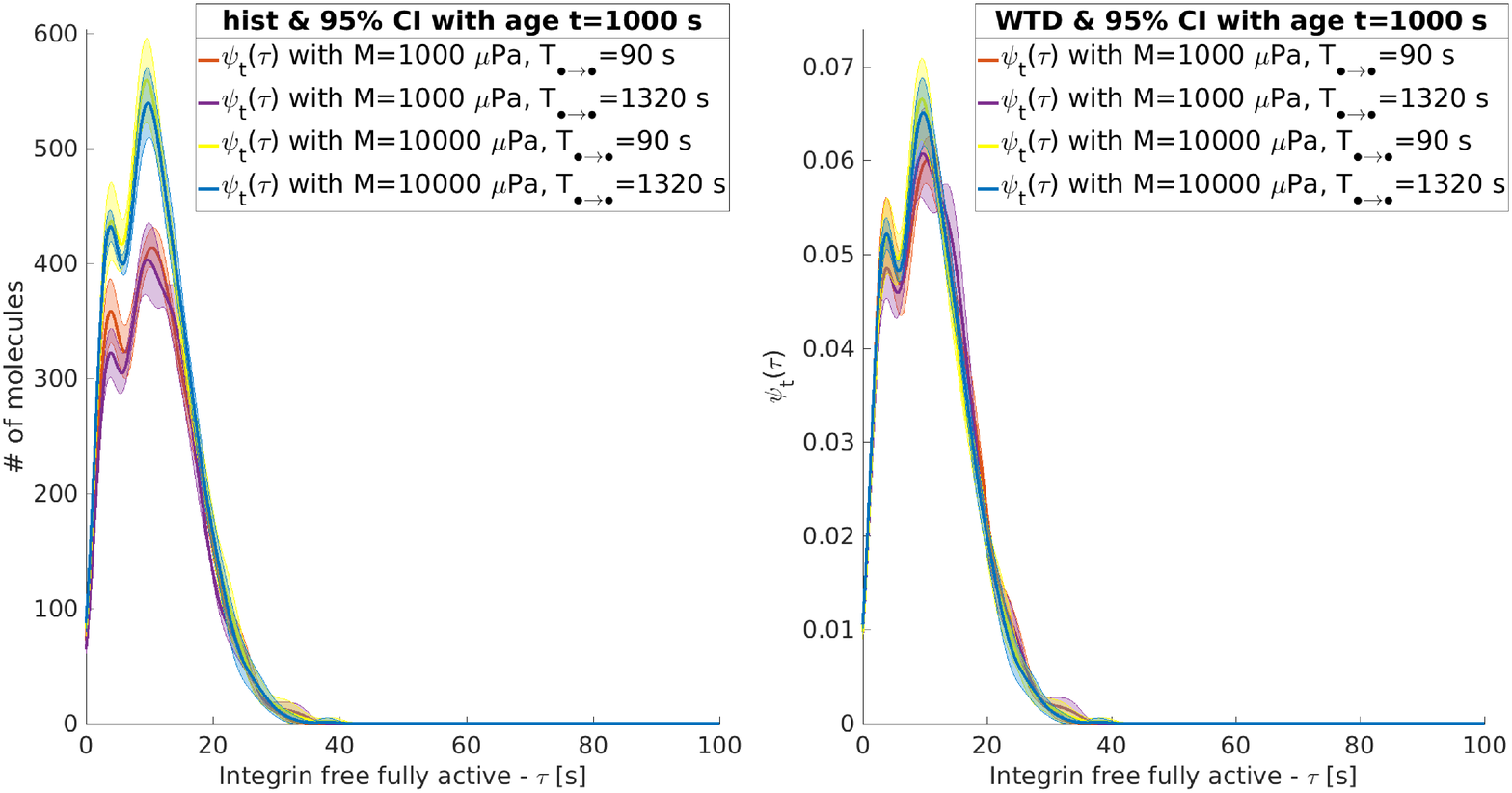}
	}
	\caption{\csentence{Active integrins.}
		The distribution of recurrence of events for active integrins is bimodal in $\tau\sim\{3, 12\}\,s$. The modes are independent from the periods and the magnitude of the mechanical load. The tail of the distributions at large $\tau$ are sensitive to the perturbation period $P$. For $P=5000\,s$, the distributions show a noisy large tail. For $P=200000\,s$, the distributions decay rapidly at zero around $tau\sim 25\,s$. At large $\tau$, the distribution does not depend on the magnitude $M$. \textbf{\protect\subref{fig_noiseA}} $\M=1000\,\mu Pa$ while $P$ and \RADP11 change; \textbf{\protect\subref{fig_noiseB}} $\M=1000\,\mu Pa$ while $P$ and \RADP12 change; \textbf{\protect\subref{fig_noiseC}}  $P=200000\,s$ while $M$ and \RADP12 change.
	}
	\label{fig_noise}
\end{figure}

\begin{figure}[h!]
	\includegraphics[trim={23ex 0ex 33ex 5ex},clip,width=12.2cm,height=7.3cm]{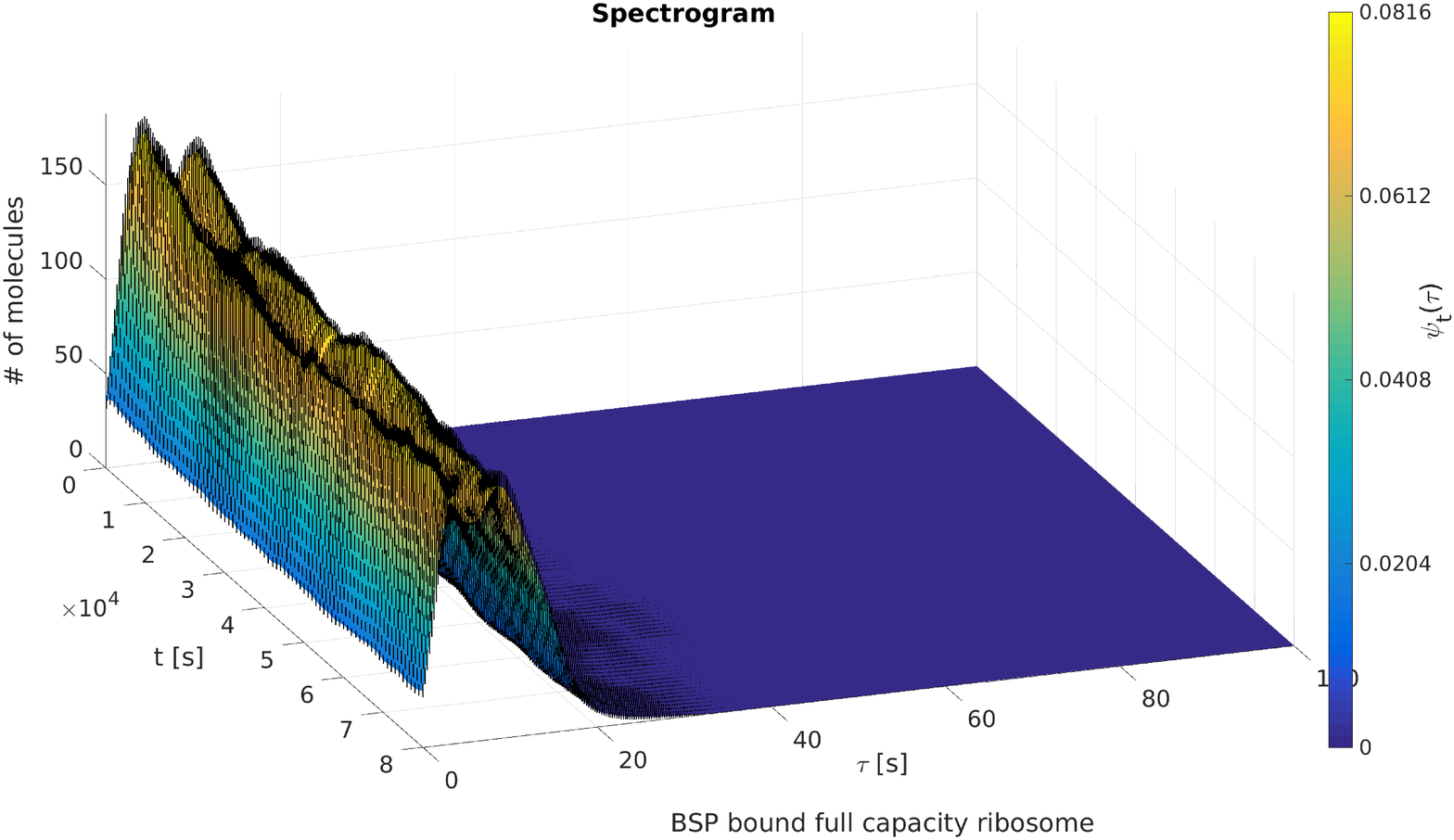}
	\caption{\csentence{BSP mRNA bound to ribosome.} 
		Spectrogram of the critical events for BSP mRNA during translation.
		On the x axis, the age $t$ of the non normalized histogram also corresponding to the beginning of an epoch; on the y axis, the intertimes $\tau$; on the z axis, the number of events registered during each epoch of length $10000\,s$. The vertical black lines crossing the surface show the 95\% CI. The color defines the value of  the WTD $\psi$  at intertime $\tau$ and  age $t$. The WTD is bimodal and the shape is constant at all $t$.
		Nonetheless, the total number of critical events decreases with the age.
	}
	\label{fig_ageing2}
\end{figure}




\clearpage

\section*{Tables}

\begin{table}[h!]
\caption{Number of molecules at time $t=0$.} \label{proportions}
      \begin{tabular}{lc}
      	\hline
      	Molecule & \# \\
      	$\s1000$ & 1  \\
      	$\s00$   & 500  \\
      	$\s01$   & 0  \\
      	$\s02$   & 0  \\
      	$\s04$   & 0  \\
      	$\s10$   & 1000  \\
      	$\s11$   & 1  \\
      	$\s12$   & 1  \\
      	$\s13$   & 1  \\
      	$\s20$   & 1000  \\
      	$\s21$   & 0  \\
      	$\s22$   & 0  \\
      	$\s23$   & 0  \\
      	$\s30$   & 32  \\
      	$\s31$   & 0  \\
      	$\s32$   & 0  \\
      	$\s33$   & 0  \\
      	$\s40$   & 3400  \\
      	$\s41$   & 0  \\
      	$\s42$   & 0  \\
      	$\s43$   & 0  \\
      	$\s50$   & 2300  \\
      	$\s51$   & 0  \\
      	$\s52$   & 0  \\
      	$\s53$   & 0  \\
      	$\s54$   & 0  \\
      	$\s60$   & 24  \\
      	$\s61$   & 0  \\
      	$\s62$   & 0  \\
      	$\s63$   & 0  \\
      	$\s70$   & 8  \\
      	$\s700$  & 0  \\
      	$\s71$   & 5 \\
      	$\s710$  & 0  \\
      	$\s711$  & 0  \\
      	$\s72$   & 5 \\
      	$\s720$  & 0  \\
      	$\s73$   & 6 \\
      	$\s730$  & 0  \\
      	$\s80$   & 0  \\
      	$\s81$   & 0  \\
      	$\s90$   & 0  \\
      	$\s91$   & 0  \\
      	$\s92$   & 0  \\
      	$\s100$  & 0  \\
      	$\s101$  & 0  \\
      	$\s102$  & 0  \\
      	$\s110$  & 0  \\
      	$\s111$  & 0  \\
      	$\s112$  & 0  \\
      	$\s120$  & 0  \\
      	$\s121$  & 0  \\
      	$\s122$  & 0  \\
      	$\s130$  & 299 \\
      	$\s131$  & 0  \\
      	$\s140$  & 301  \\
      	$\s141$  & 0  \\
      	$\s150$  & 0  \\
      	\hline
      \end{tabular}
\end{table}

\begin{table}[h!]
	\caption{Parameters' names, symbols and values.} \label{base_parameters}
	\begin{tabular}{lp{58mm}Hp{15mm}p{11mm}}
		\hline
		$\vphantom{\int^T}$ Parameter                                 &                  Symbol                   & C symb   &            Value            & \ume \\
		Mechanical min. magnitude                                     &                     m                     &          &            $100$            & $\fu$  \\
		Mechanical phase                                              &                  $\phi$                   &          &              0              &   rad   \\
		Integrin activation delay  		                              &          $\RA{\s00 \rightarrow \s01}$     & recdelay &              0              &  s   \\
		$\Int_\nact$ dissoc. time from $\Int+\FAK$ comp.              &          $\RA{\s02 \rightarrow \s00}$     & RADP1    &        $U[0,80]$              &  s   \\
		$\FAK_\act$ activation delay                                  &          $\RA{\s13 \rightarrow \s10}$     & RADPFK   &              5              &  s   \\
		$\FAK_\nact$ dissoc. time from $\Int+\FAK$ comp.              &          $\RA{\s11 \rightarrow \s13}$     & RADP2    &             30              &  s   \\
		$\Int_\act+\FAK$ dissoc. time \cm{from $\Int+\FAK+\RAS$ comp.}&          $\RA{\s12 \rightarrow \s11}$     & RADP3    &              7              &  s   \\
		$\RAS_\act$ dissoc. time from $\Int\!+\FAK\!+\RAS$ comp.  	  &          $\RA{\s21 \rightarrow \s22}$     & RADP4    &              7              &  s   \\
		$\RAS_\act$ dissoc. form $\RAS+\RAF$ comp.                    &          $\RA{\s23 \rightarrow \s22}$     & RADP5    &             14              &  s   \\
		$\RAS_\act$ relaxation time                                   &          $\RA{\s22 \rightarrow \s20}$     & RADP6    &             60              &  s   \\
		$\RAF_\act$ dissoc. time \cm{from $\RAS_\act+\RAF_\act$ comp.}&          $\RA{\s31 \rightarrow \s32}$     & RADP7    &             14              &  s   \\
		$\RAF_\act$ dissoc. time from $\RAF+\MEK$                     &          $\RA{\s33 \rightarrow \s32}$     & RADP8    &             10              &  s   \\
		Relaxation of $\RAF_\act$                                     &          $\RA{\s32 \rightarrow \s30}$     & RADP9    &             60              &  s   \\
		$\MEK_\act$ dissoc. time from $\RAF+\MEK$ comp.           	  &          $\RA{\s41 \rightarrow \s42}$     & RADP10   &             10              &  s   \\
		$\MEK_\act$ dissoc. time from $\MEK+\ERK$ comp.               &          $\RA{\s43 \rightarrow \s42}$     & RADP11   &              8              &  s   \\
		$\MEK$ relaxation time                                        &          $\RA{\s42 \rightarrow \s40}$     & RADP12   &             88              &  s   \\
		$\ERK$ activation time                                        &          $\RA{\s51 \rightarrow \s52}$     & RADP13   &              8              &  s   \\
		$\ERK$ relaxation time                                        &          $\RA{\s52 \rightarrow \s50}$     & RADP14   &             50              &  s   \\
		$\ERK_\nact$ dissoc. time from $\RUNX+\ERK$ comp.             &          $\RA{\s53 \rightarrow \s52}$     & RADP15   &             10              &  s   \\
		$\ERK_\act$ dissoc. time from $\ERK_\act+\Other$ comp.        &          $\RA{\s54\rightarrow \s52}$      & RADP16   &              6              &  s   \\
		%
	    $\RUNX_\act$ dissoc. time from $\ERK_\act+\RUNX_\nact$ comp.  &          $\RA{\s61 \rightarrow \s62}$     & RADP17   &             10              &  s   \\
		$\RUNX_\act$ dissoc. time from $\RUNX_\act+\DNA$ comp.  &          $\RA{\s63 \rightarrow \s62}$     & RADP18   &             10              &  s   \\
		$\RUNX$ relaxation time								    &          $\RA{\s62 \rightarrow \s60}$     & RADP19   &             60              &  s   \\
		$\OSXmon$ dissoc. time from $\RUNX_\act+\OSXmon$ comp.  &         $\RA{\s700 \rightarrow \s70}$     & RADP20   &             20              &  s   \\
		$\OSXmon+\DNA$ dissoc. time from $\RUNX_\act$ 		 	&         $\RA{{\s710} \rightarrow \s71}$   & RADP21   &             30              &  s   \\
		$\OSXmon+\DNA$ dissoc. time from $\AP$ 		 	&         $\RA{{\s711} \rightarrow \s71}$   & RADP22   &              0              &  s   \\
		$\OSXmul$ dissoc. time from $\RUNX_\act+\OSXmul$ comp.  &         $\RA{\s720 \rightarrow \s72}$     & RADP23   &             40              &  s   \\
		$\OSXmul+\DNA$ dissoc. time from $\RUNX_\act$ 		 	&         $\RA{{\s730} \rightarrow \s73}$   & RADP24   &             50              &  s   \\
		ribosome availability time from $\Rib_\naval+\OPN_\RNA$ &          $\RA{\s92 \rightarrow \s91}$     & RADP25   &             20              &  s   \\
		$\OPN_\RNA$ dissociation time from available ribosome				&          $\RA{\s91 \rightarrow \s90}$     & RADP26   &             75              &  s   \\
		ribosome availability time from $\Rib_\naval+\OCN_\RNA$ &         $\RA{\s102 \rightarrow \s101}$    & RADP27   &             20              &  s   \\
		$\OCN_\RNA$ dissociation time from available ribosome				&         $\RA{\s101 \rightarrow \s100}$    & RADP28   &             75              &  s   \\
		ribosome availability time from $\Rib_\naval+\ALP_\RNA$ &         $\RA{\s112 \rightarrow \s111}$    & RADP29   &             20              &  s   \\
		$\ALP_\RNA$ dissociation time from available ribosome				&         $\RA{\s111 \rightarrow \s110}$    & RADP30   &             75              &  s   \\
		ribosome availability time from $\Rib_\naval+\BSP_\RNA$ &         $\RA{\s122 \rightarrow \s121}$    & RADP31   &             20              &  s   \\
		$\BSP_\RNA$ dissociation time from available ribosome				&         $\RA{\s121 \rightarrow \s120}$    & RADP32   &             75              &  s   \\
		$\RNA$ dissociation time from $\Rib_\ncomp$       &         $\RA{\s131 \rightarrow \s130}$    & RADP33   &             100             &  s   \\
		$\RNA$ dissociation time from complete ribosome   &         $\RA{\s141 \rightarrow \s140}$    & RADP34   &             100             &  s   \\
		force tag probability                                         &             $F_\textrm{tag}$              & forcetag & $P_{\{01,02\}}=\{0.9,0.1\}$ &      \\
		$\s00$ interaction radius                                     &   $R_{\s00}$                              &          &             50              &  nm  \\ 
		$\s10$ interaction radius                                     &   $R_{\s10}$   							  &          &             50              &  nm  \\ 
		$\s20$ interaction radius                                     &   $R_{\s20}$   							  &          &             80              &  nm  \\ 
		$\s30$ interaction radius                                     &   $R_{\s30}$   							  &          &             70              &  nm  \\ 
		$\s40$ interaction radius                   				  &   $R_{\s40}$   							  &          &             50              &  nm  \\ 
		$\s50$ interaction radius                   				  &   $R_{\s50}$   							  &          &             30              &  nm  \\ 
		$\s60$ interaction radius                   				  &   $R_{\s60}$   							  &          &             30              &  nm  \\ 
		$\s70$ interaction radius                                     &   $R_{\s70}$   							  &          &             30              &  nm  \\ 
		$\s130$ interaction radius                  				  &   $R_{\s130}$  							  &          &             30              &  nm  \\ 
		cell radius                                 				  & $R_{\textrm{cell}}$  					  &          &             1000            &  nm  \\
		nucleus radius                              				  & $R_{\textrm{nucl}}$                       &          &             400             &  nm  \\
		protein' s average velocity                                   & $\overline{v}_p$                          &          &             2               &  nm/s\\
		protein' s velocity variation                                 & $\avd{v}$                                 &          &             1               &  nm/s\\
		protein' s direction variation                                & $\avd{\varphi}=\avd{\theta}$              &          &           $\pi$/10          &   rad\\
		protein translation delay    								  & $T_p$									  &protdelay &             95              &  s   \\
		transcription delay per mRNA   								  & $T_{\RNA}$								  &RNAdelay  &			  600              &  s   \\
		Integrins $\theta$ angle on cell membrane distribution        &$\Int_\theta$                              &          &          $U[0,\pi]$           &  1/rad     \\
		Integrins   $\phi$ angle on cell membrane distribution        &$\Int_\phi$                                &          &          $U[0,2\pi]$          &  1/rad    \\
		\hline

	\end{tabular}
\end{table}


\begin{table}[h!]
	\caption{Parameters ranges: Names, symbols, unit of measures and list of values simulated. Bold quantities represent the baseline values. Where no baseline is present, then all possible combinations has been considered. Each set of parameters has been independently repeated 10 times.} \label{mut_parameters}
    \begin{tabular}{lccp{11mm}}
	\hline
	Parameter & Symbol & List of values & \ume\\
	Mechanical max. magnitude & $M$ & $\{1000, 10000\}$ & $\fu$ \\
	Mechanical period & $\period$ & $\{10000, 50000, 2000000\}$ & s \\
	$\s42$ dissoc. time from $\s41$ comp. & $\RA{\s41 \rightarrow \s42}$ &$\{\mathbf{10},90,300,480,1320\}$    & s\\
	$\s42$ dissoc. time from $\s43$ comp. & $\RA{\s43 \rightarrow \s42}$ &$\{\mathbf{8},90,300,480,1320  \}$   & s\\
	$\s40$ relaxation time                & $\RA{\s42 \rightarrow \s40}$ &$\{\mathbf{60},90,300,480,1320  \}$  & s\\
	$\s52$ activation time                & $\RA{\s51 \rightarrow \s52}$ &$\{\mathbf{8},90,300,480,1320  \}$   & s\\
	$\s50$ relaxation time                & $\RA{\s52 \rightarrow \s50}$ &$\{\mathbf{90},300,480,600,1320  \}$ & s\\
	$\s52$ dissoc. time from $\s53$ comp. & $\RA{\s53 \rightarrow \s52}$ &$\{\mathbf{10},90,300,480,1320  \}$  & s\\	
	 \hline
	\end{tabular}
\end{table}

\clearpage
\newpage

{
	\makeatletter
	
	\def\makefigure@float{
		\setlength{\fboxsep}{\figure@sep}%
		\setlength{\fboxrule}{0.0\p@}%
		\fcolorbox{bmcblue}{white}{\box\bmcfloat@box}}
	
	\makeatother
	 
\begin{figure}
	\centering
\begin{tikzpicture}[->,>=stealth',shorten >=1pt,auto,node distance=1.8cm,
thick,state/.style={rectangle, rounded corners, minimum width=1cm, minimum height=1cm,,align=center,text centered,draw,font=\sffamily\normalsize\bfseries}]

\node[state] (00) {$\s00$\\00};
\node[state] (04) [right=of 00] {$\s04$\\04};
\node[state] (01) [right=of 04] {$\s01$\\01};
\node[state] (02) [right=of 01] {$\s02$\\02};

\node[state] (10) [below=of 00,yshift=-1cm] {$\s10$\\10};
\node[state] (11) [right=of 10] {$\s11$\\11};
\node[state] (12) [right=of 11] {$\s12$\\12};
\node[state] (13) [right=of 12] {$\s13$\\13};

\node[state] (20) [below=of 10,,yshift=-0.8cm] {$\s20$\\20};
\node[state] (21) [right=of 20] {$\s21$\\21};
\node[state] (22) [right=of 21] {$\s22$\\22};
\node[state] (23) [right=of 22] {$\s23$\\23};

\node[state] (30) [below=of 20] {$\s30$\\30};
\node[state] (31) [right=of 30] {$\s31$\\31};
\node[state] (32) [right=of 31] {$\s32$\\32};
\node[state] (33) [right=of 32] {$\s33$\\33};

\node[state] (40) [below=of 30,yshift=-0.5cm] {$\s40$\\40};
\node[state] (41) [right=of 40] {$\s41$\\41};
\node[state] (42) [right=of 41] {$\s42$\\42};
\node[state] (43) [right=of 42] {$\s43$\\43};

\node[state] (50) [below=of 40] {$\s50$\\50};
\node[state] (51) [right=of 50] {$\s51$\\51};
\node[state] (52) [right=of 51] {$\s52$\\52};
\node[state] (53) [right=of 52,xshift=-0.5cm] {$\s53$\\53};
\node[state] (54) [right=of 53,xshift=-1.5cm] {$\s54$\\54};

\node[state] (60) [below=of 50] {$\s60$\\60};
\node[state] (61) [right=of 60] {$\s61$\\61};
\node[state] (62) [right=of 61] {$\s62$\\62};
\node[state] (63) [right=of 62] {$\s63$\\63};

\path (02) edge [bend left,looseness=.9,opacity=0] coordinate[pos=0.1] (00b02) (00);
\node[circle,fill=black,inner sep=0pt,minimum size=5pt] (a) at (00b02) {};

\path (10) edge [bend left,looseness=.9,opacity=0] coordinate[pos=0.8] (10b11) (11);
\node[circle,fill=black,inner sep=0pt,minimum size=5pt] (a) at (10b11) {};

\path (20) edge [bend left,looseness=.9,opacity=0] coordinate[pos=0.8] (20b21) (21);
\node[circle,fill=black,inner sep=0pt,minimum size=5pt] (a) at (20b21) {};

\path (30) edge [bend left,opacity=0] coordinate[pos=0.7] (30b31) (31);
\node[circle,fill=black,inner sep=0pt,minimum size=5pt] (a) at (30b31) {};

\path (40) edge [bend left,opacity=0] coordinate[pos=0.6] (40b41) (41);
\node[circle,fill=black,inner sep=0pt,minimum size=5pt] (a) at (40b41) {};

\path (50) edge [bend left,opacity=0] coordinate[pos=0.7] (50b51) (51);
\node[circle,fill=black,inner sep=0pt,minimum size=5pt] (a) at (50b51) {};

\path (60) edge [bend left,opacity=0] coordinate[pos=0.7] (60b61) (61);
\node[circle,fill=black,inner sep=0pt,minimum size=5pt] (a) at (60b61) {};

\path[every node/.style={font=\sffamily\small}]
(00) 	edge [bend left] node [left,anchor=south] {act} (04)
(04) 	edge [bend left] node [left,anchor=south] {act} (01)
(01) 	edge [bend left] node [left,anchor=south] {R} (00)
(02) 	edge [bend left,looseness=.9] node [left,anchor= south east] {D and R p-} (00)
(04) 	edge [bend left] node [left,anchor=south] {R} (00)
(01) 	edge [bend left] node [left,anchor=south] {A} (02)

(10) 	edge [bend left] node [left,anchor=north] {A} (11)
(10) 	edge [bend left] node [left,anchor=north] {R p-} (13)
(11) 	edge [bend left] node [left,anchor=north] {A p0} (12)
(11) 	edge [bend left] node [left,anchor=north] {D and R p-} (13)
(12) 	edge [bend left] node [left,anchor=north] {D p0\/*?*/ } (11)
(13) 	edge [bend left] node [left,anchor=south] {substrate p+} (10)
(00b02)	edge [bend left] node [left,anchor=east] {D and R p-} (13)
(01) 	edge [-,dashed,bend left] node [left,anchor=south] {} (10b11)

(20) 	edge [bend left] node [left,anchor=south] {A} (21)
(21) 	edge [bend left] node [left,anchor=north] {D p+} (22)
(22) 	edge [bend left] node [left,anchor=south] {R p-} (20)
(22) 	edge [bend left] node [left,anchor=south] {A} (23)
(23) 	edge [bend left] node [left,anchor=south] {D p0\/*?*/ } (22)
(11) 	edge [-,dashed,bend left] node [left,anchor=south] {} (20b21)

(31) 	edge [bend left] node [left,anchor=north] {D p+} (32)
(33) 	edge [bend left] node [left,anchor=south east] {D p-$\phantom{a}$} (30)
(32) 	edge [bend left] node [left,anchor=south] {R p-} (30)
(30) 	edge [bend left] node [left,anchor=south] {A} (31)
(32) 	edge [bend left] node [left,anchor=south] {A} (33)
(22) 	edge [-,dashed,bend left,looseness=.4] node [left,anchor=south] {} (30b31)

(40) 	edge [bend left] node [left,anchor=south] {A} (41)
(41) 	edge [bend left] node [left,anchor=north] {D p+} (42)
(42) 	edge [bend left] node [left,anchor=south] {R p-} (40)
(42) 	edge [bend left] node [left,anchor=south] {A} (43)
(43) 	edge [bend left] node [left,anchor=south] {D p0\/*?*/ } (42)
(32) 	edge [-,dashed,bend left,looseness=1.1] node [left,anchor=south] {} (40b41)

(51) 	edge [bend left] node [left,anchor=north] {D p+} (52)
(52) 	edge [bend left] node [left,anchor=south] {A p0} (54)
(52) 	edge [bend left] node [left,anchor=south] {R p-} (50)
(53) 	edge [bend left] node [left,anchor=south] {D p0\/*?*/ } (52)
(54) 	edge [bend left] node [left,anchor=south] {D p0} (52)
(50) 	edge [bend left] node [left,anchor=south] {A} (51)
(52) 	edge [bend left] node [left,anchor=north] {A} (53)
(42) 	edge [-,dashed,bend left,looseness=0.7] node [left,anchor=south] {} (50b51)

(61) 	edge [bend left] node [left,anchor=north] {D p+} (62)
(62) 	edge [bend left] node [left,anchor=south] {R p-} (60)
(63) 	edge [bend left] node [left,anchor=south] {D} (62)
(60) 	edge [bend left] node [left,anchor=south] {A} (61)
(62) 	edge [bend left] node [left,anchor=south] {A} (63)
(52) 	edge [-,dashed,bend left,looseness=0.7] node [left,anchor=south] {} (60b61)

;
\end{tikzpicture}
\caption{Reactions in the ECM and cytoplasm. A=association, D=dissociation, R=relaxation, p+=phosphorylation, p-=dephosphorylation, p0=no phosphorylation, +1=creation, act=mechanotransduction activation, rand=chosen at random} \label{fig:S1}
\end{figure}
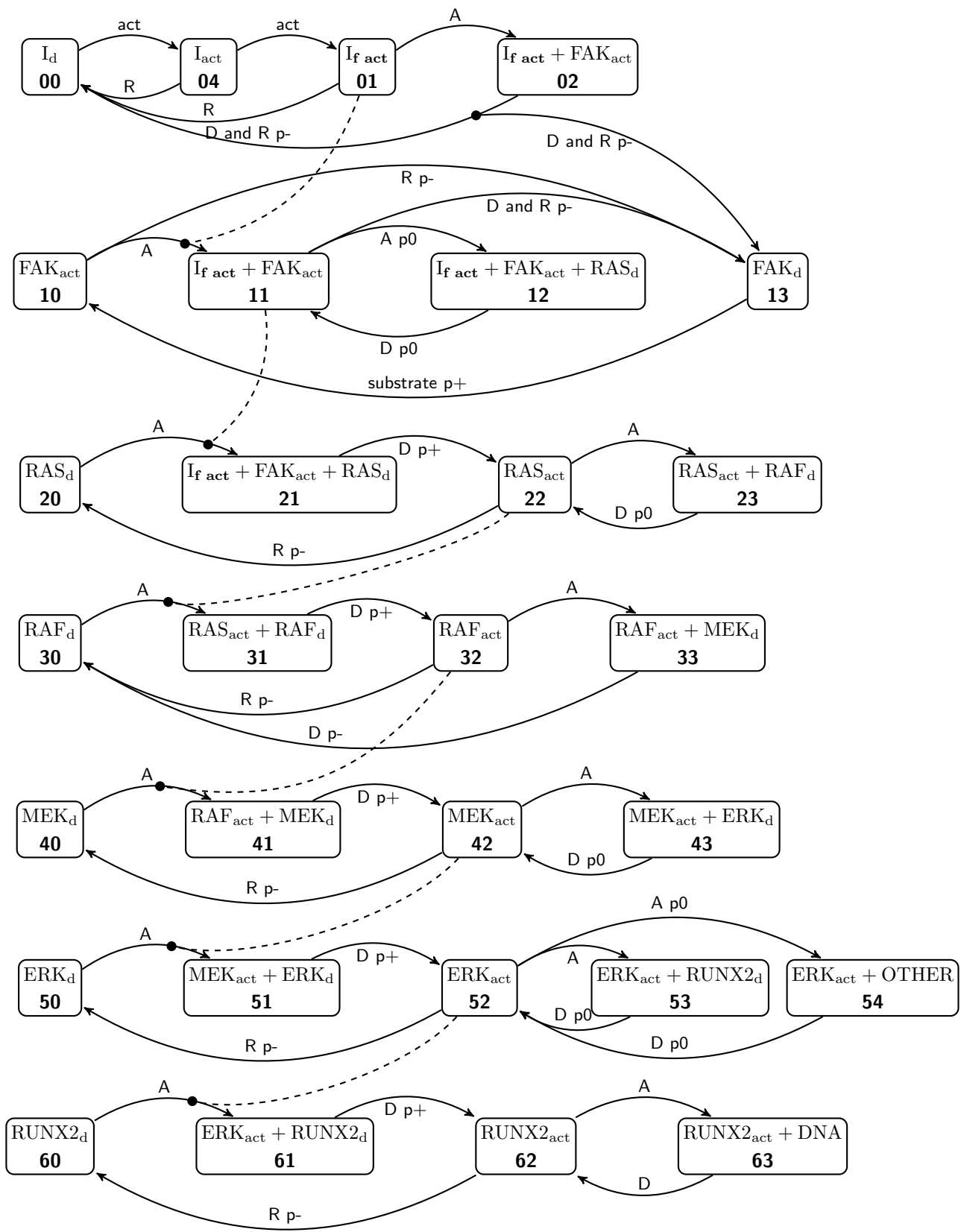

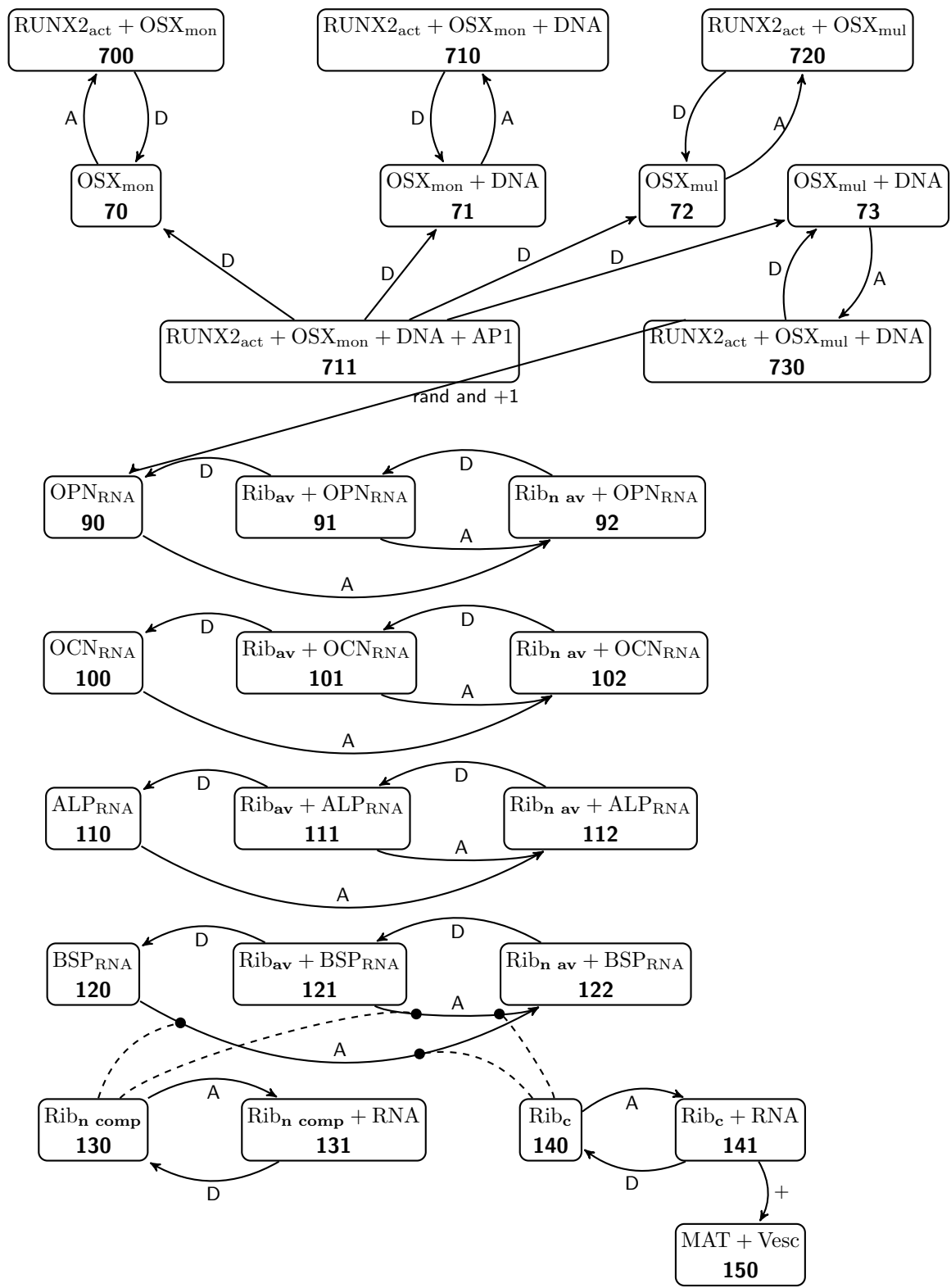
\begin{figure}
	\centering
\begin{tikzpicture}[->,>=stealth',shorten >=1pt,auto,node distance=1.5cm,
thick,state/.style={rectangle, rounded corners, minimum width=1cm, minimum height=1cm,,align=center,text centered,draw,font=\sffamily\normalsize\bfseries}]

\node[state] (700) {$\s700$\\700};
\node[state] (710) [right=of 700] {$\s710$\\710};
\node[state] (720) [right= of 710] {$\s720$\\720};

\node[state] (70) [below=of 700] {$\s70$\\70};
\node[state] (71) [below=of 710] {$\s71$\\71};
\node[state] (72) [right= of 71] {$\s72$\\72};
\node[state] (73) [right=of 72,xshift=-0.5cm] {$\s73$\\73};

\node[state] (711) [below =of 71,xshift=-2.cm] {$\s711$\\711};
\node[state] (730) [right=of 711,xshift=0.5cm] {$\s730$\\730};


\node[state] (90) [below=of 711,xshift=-4.cm] {$\s90$\\90};
\node[state] (91) [right=of 90] {$\s91$\\91};
\node[state] (92) [right=of 91] {$\s92$\\92};

\node[state] (100) [below=of 90] {$\s100$\\100};
\node[state] (101) [right=of 100] {$\s101$\\101};
\node[state] (102) [right=of 101] {$\s102$\\102};

\node[state] (110) [below=of 100] {$\s110$\\110};
\node[state] (111) [right=of 110] {$\s111$\\111};
\node[state] (112) [right=of 111] {$\s112$\\112};

\node[state] (120) [below=of 110] {$\s120$\\120};
\node[state] (121) [right=of 120] {$\s121$\\121};
\node[state] (122) [right=of 121] {$\s122$\\122};

\node[state] (130) [below=of 120] {$\s130$\\130};
\node[state] (131) [right=of 130] {$\s131$\\131};
\node[state] (140) [right=of 131] {$\s140$\\140};
\node[state] (141) [right=of 140] {$\s141$\\141};

\node[state] (150) [below=of 141,yshift=0.5cm] {$\s150$\\150};

\path (120) edge [bend right,looseness=1.0,opacity=0] coordinate[pos=0.1] (120b122a) (122);
\node[circle,fill=black,inner sep=0pt,minimum size=5pt] (a) at (120b122a) {};

\path (120) edge [bend right,looseness=1.0,opacity=0] coordinate[pos=0.7] (120b122b) (122);
\node[circle,fill=black,inner sep=0pt,minimum size=5pt] (a) at (120b122b) {};

\path (121) edge [bend right,looseness=0.4,opacity=0] coordinate[pos=0.3] (121b122a) (122);
\node[circle,fill=black,inner sep=0pt,minimum size=5pt] (a) at (121b122a) {};

\path (121) edge [bend right,looseness=0.4,opacity=0] coordinate[pos=0.7] (121b122b) (122);
\node[circle,fill=black,inner sep=0pt,minimum size=5pt] (a) at (121b122b) {};

\path[every node/.style={font=\sffamily\small}]
(700) 	edge [bend left] node [left,anchor=west] {D} (70)

(710) 	edge [bend right] node [left,anchor=east] {D} (71)

(720) 	edge [bend right] node [left,anchor=east] {D} (72)

(730) 	edge [bend left] node [left,anchor=east] {D} (73)

(711) 	edge  node [left,anchor=south] {D} (70)
(711) 	edge  node [left,anchor=east] {D} (71)
(711) 	edge  node [left,anchor=south] {D} (72)
(711) 	edge  node [left,anchor=south] {D} (73)

(70) 	edge [bend left] node [left,anchor=east] {A} (700)
(71) 	edge [bend right] node [left,anchor=west] {A} (710)
(72) 	edge [bend right] node [left,anchor=south] {A} (720)
(73) 	edge [bend left] node [left,anchor=west] {A} (730)

(730) 	edge [bend right,looseness=0.01] node [left,anchor=west] {rand and +1} (90)

(92) 	edge [bend right] node [left,anchor=north] {D} (91)
(91) 	edge [bend right] node [left,anchor=north] {D} (90)
(90) 	edge [bend right] node [left,anchor=south] {A} (92)
(91) 	edge [bend right,looseness=0.4] node [left,anchor=south] {A} (92)

(102) 	edge [bend right] node [left,anchor=north] {D} (101)
(101) 	edge [bend right] node [left,anchor=north] {D} (100)
(100) 	edge [bend right] node [left,anchor=south] {A} (102)
(101) 	edge [bend right,looseness=0.4] node [left,anchor=south] {A} (102)

(112) 	edge [bend right] node [left,anchor=north] {D} (111)
(111) 	edge [bend right] node [left,anchor=north] {D} (110)
(110) 	edge [bend right] node [left,anchor=south] {A} (112)
(111) 	edge [bend right,looseness=0.4] node [left,anchor=south] {A} (112)

(122) 	edge [bend right] node [left,anchor=north] {D} (121)
(121) 	edge [bend right] node [left,anchor=north] {D} (120)
(120) 	edge [bend right] node [left,anchor=south] {A\/*?*/ } (122)
(121) 	edge [bend right,looseness=0.4] node [left,anchor=south] {A\/*?*/ } (122)

(131) 	edge [bend left] node [left,anchor=north] {D} (130)
(141) 	edge [bend left] node [left,anchor=north] {D} (140)

(130) 	edge [bend left] node [left,anchor=north] {A} (131)
(140) 	edge [bend left] node [left,anchor=north] {A} (141)

(130) 	edge [-,dashed,bend left] node [left,anchor=north] {} (120b122a)
(130) 	edge [-,dashed,bend left,looseness=0.4] node [left,anchor=north] {} (121b122a)
(140) 	edge [-,dashed,bend right] node [left,anchor=north] {} (120b122b)
(140) 	edge [-,dashed,bend right,looseness=0.3] node [left,anchor=north] {} (121b122b)
(141) 	edge [bend left] node [left,anchor=west] {+} (150)
;
\end{tikzpicture}
\caption{Reactions in the nucleus. A=association, D=dissociation, R=relaxation, p+=phosphorylation, p-=dephosphorylation, p0=no phosphorylation, +1=creation, rand=chosen at random} \label{fig:S2}
\end{figure}

}




\end{backmatter}
\end{document}